\documentclass[fleqn,usenatbib]{mnras}

\usepackage{mathptmx}

\usepackage[T1]{fontenc}
\usepackage{ae,aecompl}

\usepackage{graphicx}	
\usepackage{amsmath}	
\usepackage{amssymb}	
\usepackage{bm}         
\usepackage{nth}
\usepackage{etoolbox}
\usepackage[normalem]{ulem}
\usepackage{color}

\usepackage{pdflscape}

\makeatletter
\patchcmd\@combinedblfloats{\box\@outputbox}{\unvbox\@outputbox}{}{%
  \errmessage{\noexpand\@combinedblfloats could not be patched}%
}%
\makeatother

\newcommand{\RC}{R_\mathrm{C}}
\newcommand{\Clim}{$C_{\,\mathrm{lim}}$}
\newcommand{\Plim}{$P_{\,\mathrm{lim}}$}
\newcommand{\Proto}{\mbox{Proto}}
\newcommand{\PProto}{\mbox{PartProto}}
\newcommand{\ProtoF}{\mbox{ProtoField}}
\newcommand{\Field}{\mbox{Field}}
\newcommand{\SelNine}{S$_{\mathrm{MAS}9}$}
\newcommand{\SelTen}{S$_{\mathrm{MAS}10}$}
\newcommand{\SelSFR}{S$_{\mathrm{SFR}1}$}
\newcommand{\SelSFRFive}{S$_{\mathrm{SFR}5}$}

\newcommand{\App}[1]{Appendix~\ref{sec:#1}}

\newcommand{\Fig}[1]{Figure~\ref{fig:#1}}
\newcommand{\Sec}[1]{Section~\ref{sec:#1}}
\newcommand{\Tab}[1]{Table~\ref{tab:#1}}

\title[Characterising and Identifying Protoclusters]{Characterising and Identifying Galaxy Protoclusters}

\author[C. C. Lovell et al.]{
Christopher C. Lovell,$^{1}$\thanks{E-mail: c.lovell@sussex.ac.uk}
Peter  A. Thomas,$^{1}$
and Stephen M. Wilkins$^{1}$
\\
$^{1}$Astronomy Centre, Department of Physics and Astronomy, University of Sussex, Brighton, BN1 9QH, UK
}

\date{Accepted XXX. Received YYY; in original form ZZZ}

\pubyear{2017}

\begin{document}
\label{firstpage}
\pagerange{\pageref{firstpage}--\pageref{lastpage}}
\maketitle
\begin{abstract}
We study the characteristics of galaxy protoclusters using the latest \textsc{L-galaxies} semi-analytic model.
Searching for protoclusters on a scale of $\sim 10 \, \mathrm{cMpc}$ gives an excellent compromise between the completeness and purity of their galaxy populations, leads to high distinction from the field in overdensity space, and allows accurate determination of the descendant cluster mass.
This scale is valid over a range of redshifts and selection criteria.
We present a procedure for estimating, given a measured galaxy overdensity, the protocluster probability and its descendant cluster mass for a range of modelling assumptions, particularly taking into account the shape of the measurement aperture.
This procedure produces lower protocluster probabilities compared to previous estimates using fixed size apertures.
The relationship between AGN and protoclusters is also investigated, and shows significant evolution with redshift; at $z \sim 2$ the fraction of protoclusters traced by AGN is high, but the fraction of all AGN in protoclusters is low, whereas at $z \geqslant 5$ the fraction of protoclusters containing AGN is low, but most AGN are in protoclusters.
We also find indirect evidence for the emergence of a passive sequence in protoclusters at $z \sim 2$, and note that a significant fraction of all galaxies reside in protoclusters at $z \geqslant 2$, particularly the most massive.
\end{abstract}

\begin{keywords}
galaxies: clusters: general -- galaxies: high redshift -- galaxies: statistics
\end{keywords}


\section{Introduction}

Present day galaxy clusters are the most massive collapsed objects in the Universe.
Each is composed of a virialised dark matter halo ($> 10^{14} \, M_{\odot}$) hosting hundreds of galaxies that exhibit a clear red sequence \citep{dressler_galaxy_1980, vikhlinin_clusters_2014}.
The progenitors of clusters are known as \textit{protoclusters}, commonly defined as the ensemble of objects that will end up in the cluster at $z = 0$ \citep{muldrew_what_2015, overzier_realm_2016}.
They tend to be highly spread out spatially, up to 20\,cMpc across by $z \sim 2$ and greater at higher redshifts \citep{suwa_protoclusters_2006, chiang_ancient_2013, muldrew_what_2015}, and host accelerated galaxy growth; approximately 50\% of the stars that end up in the brightest cluster galaxy are formed by $z \sim 5$ \citep{lucia_hierarchical_2007}.
Clusters assemble late, the most massive reaching a mass of $10^{14} \, M_{\odot}$ at $z \sim 2$, and the majority assemble half their mass by $z \sim 0.6$ \citep{wu_rhapsody._2013}; protoclusters are more spread out and diffuse distributions of matter at high redshifts, rather than a single, massive, virialised halo, and hence do not exhibit observational characteristics of present day clusters, such as thermal X-ray emission from a hot Intra-Cluster Medium (ICM) \citep{overzier_realm_2016}.
They do, however, necessarily trace overdensities of matter in the early Universe \citep{angulo_journey_2012}, which manifest as visible overdensities in the galaxy population.
This is the primary means of discriminating protoclusters from the field, and the magnitude of the overdensity is positively correlated with the descendant cluster mass \citep{overzier_cdm_2009, chiang_ancient_2013, orsi_environments_2016}.

Observational searches for protoclusters tend to adopt one of two approaches: `blind' searches for surface overdensities of galaxies, and focused observations around biased tracers.
The former typically work by identification of surface overdensities in wide field photometric surveys of Lyman break galaxies (LBGs) and narrow band imaging of emission line galaxies \citep{shimasaku_subaru_2003, adams_hetdex_2011, spitler_first_2012, chiang_discovery_2014, shimakawa_2018}, which are often followed up and confirmed spectroscopically \citep{toshikawa_discovery_2012, diener_protocluster_2015, toshikawa_systematic_2016}.
The VIMOS Ultra Deep Survey, the largest purely spectroscopic search, recently announced the discovery of a massive candidate at $z \sim 4.57$ \citep[VUDS, ][]{fevre_vimos_2015, lemaux_vimos_2017}.

The second method takes advantage of objects thought to represent biased tracers of the underlying matter distribution, such as dusty star forming galaxies \citep{capak_massive_2011, casey_dusty_2014}, Ly-$\alpha$ emitting blobs or extended Ly-$\alpha$ absorbers \citep{hennawi_quasar_2015, Cai_mapping_2016}, High-redshift Radio Galaxies (HzRGs) and quasars.
Using biased tracers to search for protoclusters is cheaper than performing wide, deep surveys. However, the uncertainty in their correlation could arguably make them unreliable: they may not probe a significant fraction of protoclusters \citep{orsi_environments_2016}, or produce an unrepresentative sample of the population.

A significant number of protoclusters have been found targeting HzRGs \citep{fevre_clustering_1996, miley_spiderweb_2006, venemans_protoclusters_2007, galametz_galaxy_2010, hatch_galaxy_2011, koyama_massive_2012, wylezalek_galaxy_2013, shimakawa_identification_2014, cooke_z_2014}.
Both \cite{ramos_almeida_environments_2013} and \cite{hatch_why_2014} propose that the large-scale overdense environment may be causally connected to the presence of a radio-loud AGN, which may not necessarily reside at the peak of the overdensity.
Searches surrounding quasars, on the other hand, have turned up a less conclusive picture; whilst many luminous quasars are clearly located in overdensities \citep{husband_are_2013, adams_discovery_2015, hennawi_quasar_2015, morselli_primordial_2014, mazzucchelli_no_2017, miller_investigating_2016}, many reside in average overdensity environments \citep{willott_imaging_2005, uchiyama_luminous_2017}.

Given a galaxy overdensity measured with one of the above approaches, we wish to know the probability that it represents a protocluster, and an estimate of its descendant cluster mass, a useful property on which many other protocluster properties (size, maturity) depend.
They can be estimated analytically \citep[][e.g.]{steidel_large_1998}, or from cosmological simulations \citep{suwa_protoclusters_2006}: protocluster probability is typically estimated by taking the ratio of regions with a given overdensity that evolve in to protoclusters to those that do not \citep{chiang_ancient_2013, chiang_discovery_2014}, and estimates of descendant mass have been inferred empirically from the typical descendant mass of a protocluster with similar overdensity \citep{orsi_environments_2016}.
Approaches such as these have been used in the construction of some of the first protocluster catalogues \citep{franck_candidate_2016, franck_candidate_2016-1}.

Measures of overdensity are typically carried out with apertures or nearest neighbour approaches, the former showing greater correspondence with the actual 3D overdensity \citep{shattow_measures_2013}, though orientation, aperture size and redshift uncertainty can have a significant effect on the quantitative overdensity value \citep{chiang_ancient_2013, monaco_tracing_2005}, which can in turn affect probability and mass estimates.
In particular, redshift uncertainty acts to effectively elongate the measurement aperture, which lowers the measured overdensity by including more field volume.
It also complicates the definition of a protocluster in simulations - when does a randomly selected irregular aperture represent a protocluster or not?
Prior to virialisation, protoclusters are an integral part of the high redshift cosmic web, tracing the nodes and filaments of the large scale structure \citep{overzier_realm_2016}, which also complicates their identification and discrimination from the field, particularly so when using elongated apertures due to the risk of alignment.

In this paper we present an improved procedure for generating descriptive statistics of protoclusters that models the shape of the measurement aperture, and a robust protocluster definition for generating probabilities.
We also investigate the spatial characteristics of protoclusters in order to determine whether the simplifying assumption of spherical symmetry is accurate, and how best to discriminate protoclusters from the field.

We use the halo catalogues from the publicly available Millennium Simulation, scaled to the \textsf{Planck1} cosmology\footnote{$\Omega_{0} = 0.315$, $\Omega_{\Lambda} = 0.685$, $h = 0.673$, $n_{s} = 0.961$ and $\sigma_{8} =	0.826$ \citep{planck_collaboration_planck_2014}}, coupled with the latest \textsc{L-Galaxies} semi-analytic model \citep{henriques_galaxy_2015} to populate our halos with galaxies and predict their nuclear properties.
The large size of Millennium allows us to study the progenitors of a sufficient number of high mass clusters to produce usable statistics on the protocluster population.
Our focus is on $z \geqslant 2$, where protoclusters are on the whole unvirialised and difficult to identify using typical cluster finding techniques.
We do not model the photometric properties of galaxies to avoid introducing further assumptions.

We describe our definitions, selection criteria and the \textsc{L-galaxies} model in \Sec{method}, the galaxy population in protoclusters as a whole (\Sec{population}), then characterise protoclusters in terms of their shapes (\Sec{triaxial}) and sizes (\Sec{spherical}).
\Sec{agn} investigates the relationship between protoclusters and AGN, and finally in \Sec{gods} we outline a procedure for generating improved statistics on galaxy overdensities, and apply our procedure to candidates from the literature (\Sec{discussion}).

\section{Models and Methods}
\label{sec:method}

\subsection{Simulation}
\label{sec:sims}

We use the Millennium dark matter $N$-body simulation \citep{springel_simulations_2005}, which evolves $2160^{3}$ particles (with mass $1.43 \times 10^9 \, M_{\odot}$) from $z = 127$ to $z=0$, in a comoving box with side length $480.3\; h^{-1}\,\mathrm{cMpc}$.
The original simulation was run using \textsf{WMAP1} cosmological parameters \citep{spergel_first-year_2003}, however in this paper we use the halo properties rescaled to the \textsf{Planck1} cosmology using the method described in \cite{angulo_one_2010}.

\textsc{L-galaxies}, or the Munich SAM, is a Semi-Analytic Model of galaxy evolution.
The latest version \citep{henriques_galaxy_2015} is an update to that presented in \cite{guo_dwarf_2011} that uses the Planck first year cosmological parameters, and better predicts the abundance of low mass galaxies at $z \geq 1$.
Using the abundance and passive fractions of galaxies at $z \leqslant 3$ the SAM model parameters are constrained using an MCMC approach, which reproduces key observables during this epoch such as the stellar mass and luminosity function.
The results have been tested and compared against various properties of the galaxy population and found to be in good agreement.
Despite being tuned to low redshift observables, the model also shows good agreement with high redshift galaxy properties, such as the stellar mass and luminosity function, out to $z = 7$ \citep{clay_galaxy_2015}. 
A full description of the model is provided in the appendix to \cite{henriques_galaxy_2015}.

The growth of supermassive black holes is modelled in \textsc{L-Galaxies} through two mechanisms \citep{croton_many_2006, henriques_galaxy_2015}.
The first, labelled \textit{quasar mode} growth, is triggered by a galaxy merger.
The black holes merge instantaneously, and are then fed cold gas driven toward the nuclear region of the galaxy by turbulent motions induced by the merger.
The second mechanism, labelled \textit{radio mode} growth, is fed by hot gas from the halo, and leads to the formation of hot bubbles and jets.
The quasar mode is the most effective mechanism by which black holes grow in the model, though the accretion is not explicitly associated with any feedback, except through supernovae feedback associated with the post-merger starburst in the case of a gas rich merger.
In contrast, radio mode feedback leads to negligible black hole growth but produces efficient feedback that prevents the infall of cold gas in the largest halos.

The \textsc{L-galaxies} AGN model is a relatively simple phenomenological representation of the physical processes that lead to observable quasar and radio activity.
It does not, for example, provide spin information, necessary for a complete description of the radio jet power \citep{fanidakis_grand_2011}.
As such, it does not match quantitative observational constraints on the accretion rate and black hole mass at high redshift.
However, in this study we are primarily interested in the number density and spatial distribution of AGN and their hosts with regards to protoclusters; since AGN activity in the model depends explicitly on host halo mass, and implicitly on environment, a simple accretion cut should allow us to evaluate their coincidence with overdensities at high-$z$.
A detailed description of AGN number densities, host halo masses and selection criteria is described in \Sec{agn}.

\subsection{Definitions}
\label{sec:term}

We define as a \textit{cluster} any Friends-of-Friends (FoF) halo at $z = 0$ with $M_{200} / M_{\odot} > 10^{14}$.
Using this definition we identify 3825 clusters.
We treat everything within $R_{200}$ of the halo centre as a cluster member, and anything outside a cluster is labelled part of the \textit{field}.

Throughout the paper, we use the following definition of a protocluster: that it is the ensemble of all objects that eventually end up in a present day cluster.
Specifically, a protocluster member is any halo or galaxy whose descendant at $z=0$ lies within $R_{200}$ of a cluster.
To identify the protoclusters at a given epoch we follow the merger tree rooted on each subhalo in the cluster at $z=0$, including the central subhalo, back in time to identify all progenitor halos and their galaxies.

\subsection{Galaxy selection}
\label{sec:selection}

We apply four galaxy selection criteria, identical to those employed in \cite{chiang_ancient_2013}, with an additional high star formation rate selection, at snapshots corresponding approximately to $z = [2, 3, 4, 5, 6, 7, 8, 9.5]$:
\begin{align}
\mathrm{S_{MAS9}}&: &\mathrm{log_{10}}(M_{*} / \, M_{\odot}) &> 9 \\
\mathrm{S_{MAS10}}&: &\mathrm{log_{10}}(M_{*} / \, M_{\odot}) &> 10 \\
\mathrm{S_{SFR1}}&: &\mathrm{SFR} / (M_{\odot} \, \mathrm{yr}^{-1}) &> 1 \\
\mathrm{S_{SFR5}}&: &\mathrm{SFR} / (M_{\odot} \, \mathrm{yr}^{-1}) &> 5 \,.
\end{align}
The star formation rate selections (\SelSFR\ and \SelSFRFive) most closely resemble the selection of line emission galaxies using narrow-band filters \citep[e.g.][]{cooke_z_2014}.

\subsection{Overdensity}
\label{sec:od}
Measures of protocluster overdensity using fixed volume apertures lead to greater consistency with redshift and better correspondence with the true 3D overdensity as compared to nearest neighbour approaches \citep{muldrew_measures_2012, shattow_measures_2013}.
We define overdensity as
\begin{equation}
\delta_{g}(\bm{x},V,z) \,\equiv\, \frac{n_{g}(\bm{x},V,z)}{\langle n_{g}(V,z) \rangle} \; - \; 1 \,,
\end{equation}
where $\delta_{g}(\bm{x},V,z)$ is the overdensity within a volume $V$ centred on position $\bm{x}$ at redshift $z$.
The volume can be spherical, $V = \frac{4}{3}\, \pi\, R^{3}$, or cylindrical, $V = \pi \,R^{2}\,D$, where $R$ is the radius on the plane of the sky and $D$ is the depth in the line-of-sight direction; we make clear in the relevant sections which volume is being used.
$n_{g}(\bm{x},V,z)$ is the number of selected galaxies within the chosen volume centred on $\bm{x}$, and $\langle n_{g}(V,z) \rangle$ is the mean number of selected galaxies in a volume of this size averaged over the entire simulation.

Where we wish to compared measured overdensities as closely as possible to observations, we must take into account peculiar motions along the Line-of-Sight (LoS). High velocities along the LoS could move a galaxy into or out of a protocluster region, boosting or diminishing the measured overdensity, respectively.
To account for this effect, we transform the LoS coordinate as follows:
\begin{equation}
d' = d + \frac{v_\mathrm{\, los}}{a(z) \; H(z)} \,.
\end{equation}
Here $d$ is the original comoving coordinate value, $d'$ is the transformed coordinate, $v_\mathrm{los}$ is the peculiar galaxy velocity in the LoS direction, $a$ is the expansion factor and $H(z)$ is the Hubble parameter at redshift $z$.

\section{Results}
\label{sec:results}
\subsection{The Protocluster Galaxy Population}
\label{sec:population}

\begin{figure}
	\includegraphics[width=\columnwidth]{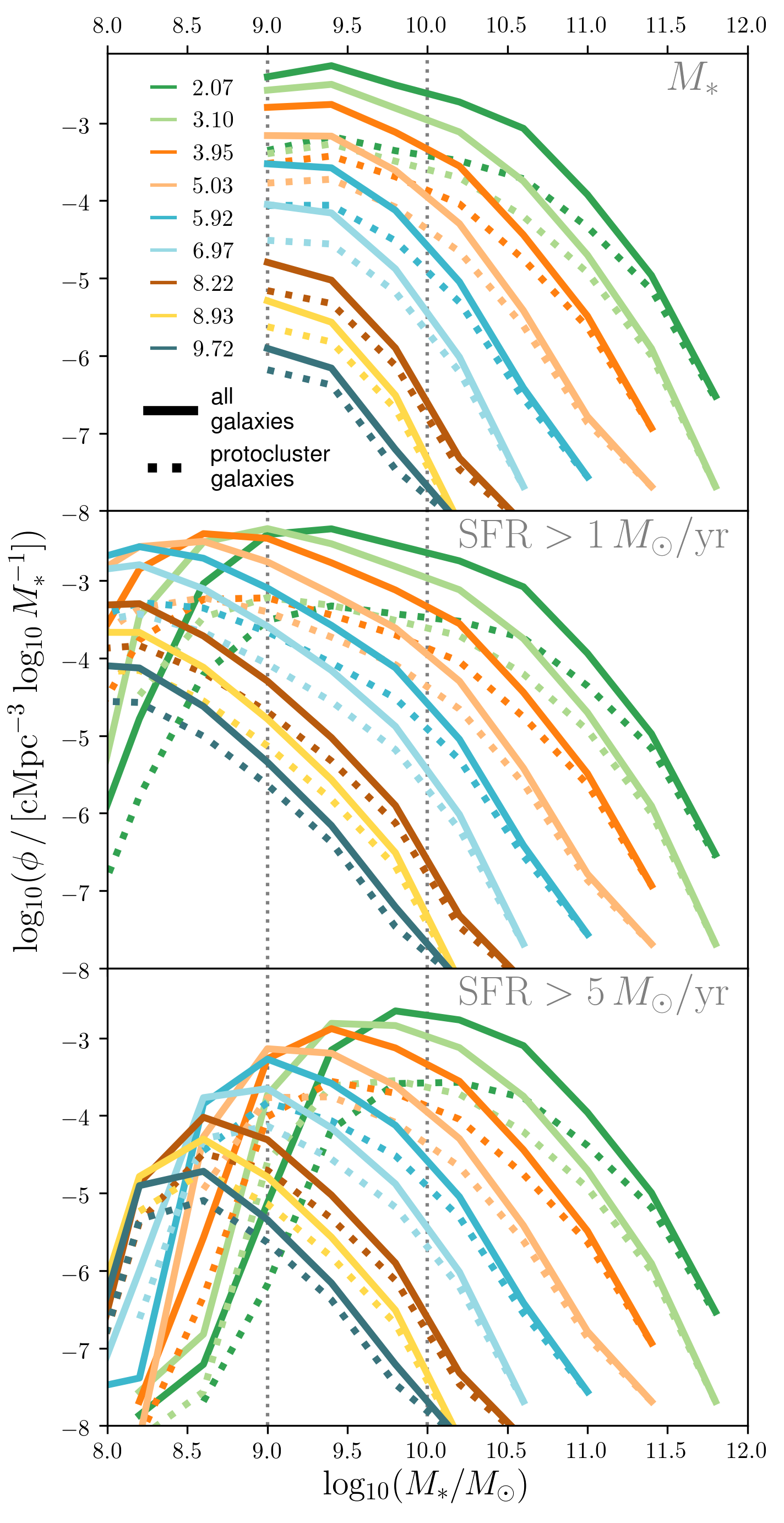}
    \caption{GSMF for all selections. The vertical dotted lines delimit the \SelNine\ and \SelTen\ selections. Solid lines show the full galaxy population, dashed lines show galaxies in protoclusters. The highest mass galaxies preferentially appear in protocluster environments, and there is a dearth of low mass galaxies evidenced by the flat low mass slope, as seen in \protect\cite{muldrew_what_2015} for a previous version of the model. \SelSFR\ extends to lower stellar masses, but has little effect on the high mass end. \SelSFRFive\ truncates the selection of low mass galaxies, though the shape of the high mass slope is again unaffected.}
    \label{fig:gsmf}
\end{figure}

\begin{figure}
	\includegraphics[width=\columnwidth]{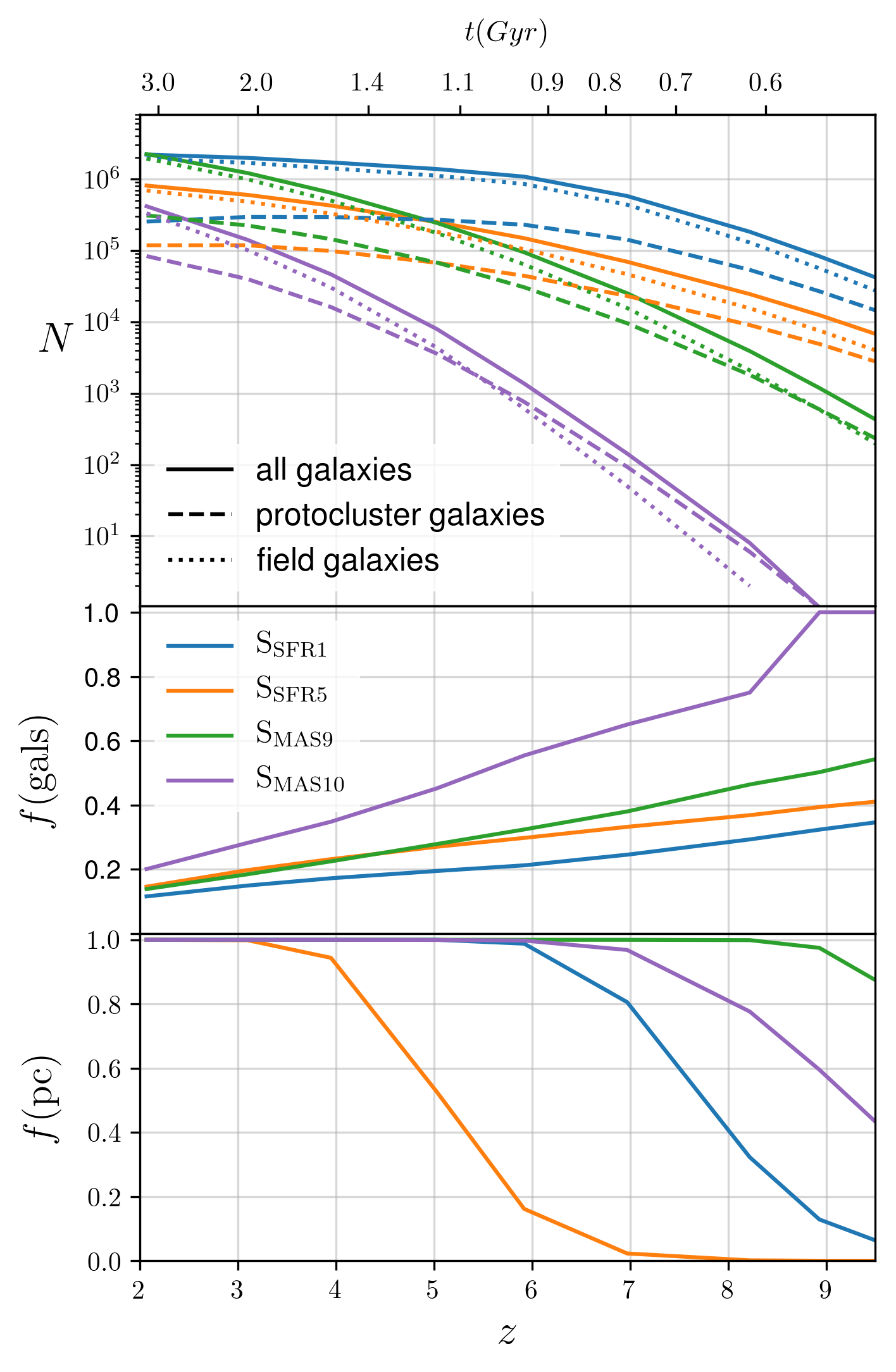}
   \caption{\textit{Top:} Number of galaxies over time, for all galaxies (solid), protocluster galaxies (dashed) and field galaxies (dotted), for each selection. \textit{Middle:} The fraction of galaxies in each selection that reside in protoclusters. \textit{Bottom:} The fraction of protoclusters that contain at least one galaxy in the given selection.}
   \label{fig:pc_fractions}
\end{figure}

We begin by looking at the evolution of the galaxy population as a whole from $2 \leq z \leq 9$ divided into protocluster and field designations.
\Fig{gsmf} shows the Galaxy Stellar Mass Function (GSMF) for each selection criteria at each redshift, along with the biased GSMF for those galaxies that reside in protoclusters.
The most massive galaxies are more likely to reside in protoclusters, and there is a dearth of low mass galaxies in protoclusters compared to the field, similar to trends seen in protocluster observations \citep{steidel_spectroscopic_2005, strazzullo_galaxy_2013, cooke_z_2014}.
The normalisation is significantly lower at the intermediate to low mass range, similar to that seen in the $z < 1$ cluster environment \citep{vulcani_galaxy_2011}.

The top panel of \Fig{pc_fractions} shows the number of galaxies over cosmic time, split into field and protocluster populations.
The number of star forming (\SelSFR\ \& \SelSFRFive) galaxies in protoclusters plateaus at $z \sim 5$, whilst similarly star forming galaxies continue to increase in number in the field.
The middle panel shows the fraction of all galaxies from each selection that reside in protoclusters; at $z=2$ a minority (10-20 \%) of galaxies lie in protoclusters, rising to $\frac{1}{4}$, $\frac{1}{3}$, $\frac{1}{2}$ and 1 at $z > 9$ for \SelSFR, \SelSFRFive, \SelNine\ and \SelTen, respectively.
Conversely, the bottom panel of \Fig{pc_fractions} shows the fraction of protoclusters that contain \textit{at least one} galaxy from each selection; all protoclusters contain at least a \SelNine\ mass galaxy up to the most extreme redshifts, whereas \SelSFRFive\ galaxies are only observed in a majority of protoclusters at $z < 5$.

For \SelTen\ galaxies at $z > 6$ there is a $> 50\%$ chance they reside in a protocluster, and $> 50\%$ of all protoclusters contain at least one \SelTen\ galaxy up to extreme redshifts; such galaxies can act as beacons of protocluster regions solely by virtue of their existence.

\subsection{Triaxial Modelling}
\label{sec:triaxial}

\begin{figure*}
	\includegraphics[width=\textwidth]{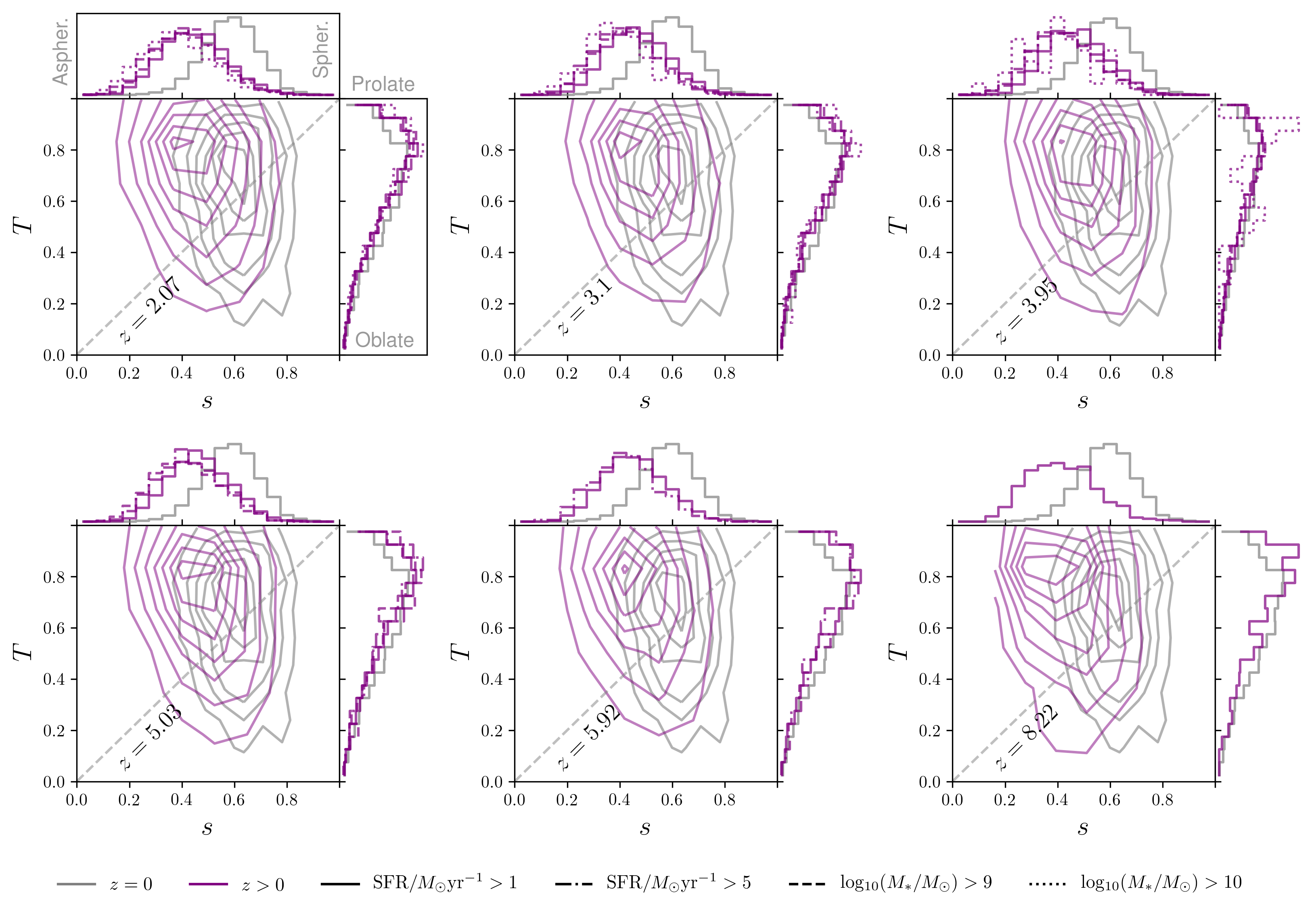}
    \caption{$s$ ratio (a measure of sphericity) and $T$ parameter (a measure of the form of apshericity) distributions. Each panel shows the 2D (for \SelSFR) and marginal (selection labelled) distributions at a given redshift. Values of $s$ close to 1 indicate spherical distributions, values close to 0 aspherical. Values of $T$ close to 1 indicate prolate distributions, values close to 0 oblate; if the $s$ distribution suggests a spherical distribution then the nature of the asphericity is unimportant. Protoclusters tend to be aspherical, with a prolate distribution, and this asphericity is pronounced at high redshift. The $z=0$ distributions (for \SelNine, since there are an insufficient number of galaxies with high star formation rates at high-$z$) are shown in grey for comparison.}
    \label{fig:sT_2d}
\end{figure*}

We have seen that protocluster galaxy membership evolves significantly with redshift and depends critically on the selection.
We now look at the distribution of galaxies within protoclusters, and present the first model of protocluster shapes, a simple triaxial model of the galaxy spatial distribution at high redshift, in order to determine the extent to which they differ from the simplifying assumption of spherical symmetry.
We acknowledge that such a simple model cannot probe collapsed structure such as groups and filaments within the protocluster, but it is capable of tracing the most prominent structure (if it exists), and provides insight into the global spatial asphericity, important for overdensity measurements.

The length and direction of each semi-axis in the triaxial model can be derived from the eigenvalues and eigenvectors, respectively, of the inertia tensor of the galaxy distribution.
The components of the inertia tensor are given by
\begin{equation}
\bm{I}_{ij} = \sum_{n=1}^{N_{g}} ( \bm{r}_{n}^{\,2}\, \delta_{ij} - r_{n,i} r_{n,j}) \,,
\end{equation}
where $N_{g}$ is the number of galaxies in the protocluster, $\bm{r}_{n}$ is the position vector of the $n^{\,\mathrm{th}}$ galaxy, and $i$ and $j$ are the tensor indices ($i, j \in {1, 2, 3}$).
We ignore the full matter distribution and focus on observable tracers, setting all galaxies to have equal mass, and also ignore redshift space distortions, so that any asphericity is randomly orientated.
The moments of inertia of $\bm{I}$ are given by its eigenvalues, $\lambda_{1} \geqslant \lambda_{2} \geqslant \lambda_{3}$, which can be translated into the relative axis lengths ($a \geqslant b \geqslant c$):
\begin{align}
a &= \sqrt{\frac{5}{2N_{g}} (\lambda_{2} + \lambda_{3} - \lambda_{1})} \\
b &= \sqrt{\frac{5}{2N_{g}} (\lambda_{1} + \lambda_{3} - \lambda_{2})} \\
c &= \sqrt{\frac{5}{2N_{g}} (\lambda_{1} + \lambda_{2} - \lambda_{3})} \,,
\end{align}
Using these axis lengths we introduce three axis ratios,
\begin{equation}
s \equiv \frac{c}{a},\ \ \  q \equiv \frac{b}{a},\ \ \  p \equiv \frac{c}{b} \,.
\end{equation}
Of these, $s$ is of particular value as a measure of sphericity: where $s = 1$, the distribution is spherical, and where $s \sim 0$, the distribution is highly aspherical.
The $q$ and $p$ ratios can be used together to deduce the form of the asphericity: where $q \sim 1(0)$ the distribution is oblate (prolate), and where $p = 1(0)$ the distribution is prolate (oblate).
An alternative measure of the form of the asphericity is the Triaxiality parameter \citep{franx_ordered_1991},
\begin{equation}
T = \frac{a^{2} - b^{2}}{a^{2} - c^{2}}
\end{equation}
which measures whether an ellipsoid is prolate ($T = 1$) or oblate ($T = 0$), but does not measure the degree of asphericity.

Similar shape analysis has been applied to a range of astrophysical objects, including the profiles of cluster dark matter halos \citep{thomas_structure_1998, wu_rhapsody._2013}.
In such cases the reduced inertia tensor, which weights particles near the centre of the halo more highly, is often used \citep{schneider_shapes_2012}.
Since protocluster profiles are less centrally concentrated than clusters (it is often difficult to unambiguously identify the protocluster centre), and are more likely to contain multiple filamentary structures, we use the unweighted inertia tensor to characterise the entire shape.
\cite{bett_spin_2007} also note that particle discreteness can affect the determination of shape parameters using the inertia tensor; to mitigate this effect we ignore those selections where the average number of tracer galaxies in a protocluster drops below 20 at a given redshift.

\Fig{sT_2d} shows the combined and marginal distributions of the $s$ ratio and $T$ parameter at different redshifts\footnote{There is significant evolution in the number of galaxies in protoclusters selected by stellar mass or star formation rate throughout cosmic time, necessitating comparison between selections where there are insufficient galaxies to make a robust shape measurement: for example, galaxies at $z=0$ are selected using the \SelNine\ criteria, since there are not enough galaxies with high star formation rates at late times, and at $z \geq 2$ only \SelSFR\ is shown for the combined distribution, as it is the most populous selection.}.
At $z=0$ (shown in grey) the majority of clusters, as traced by their galaxies, are mildly aspherical with a prolate configuration.\footnote{At $z=0$, $\bar{s} = 0.61$, $\sigma_{s}=0.10$, $\bar{T} = 0.61$, $\sigma_{\,T} = 0.20$. This asphericity is greater than that measured using the full dark matter particle information \citep{schneider_shapes_2012}.}
Protoclusters, in comparison, are more aspherical, and the majority are prolate\footnote{At $z=3$, $\bar{s}=0.50$, $\sigma_{s}=0.12$, $\bar{T}= 0.65$ and $\sigma_{T}= 0.20$.}.

The \SelNine\ and \SelTen\ selections (shown in the marginal distributions of \Fig{sT_2d}) exhibit greater asphericity than those selected by star formation rate: those tracer galaxies that make the selection cut tend to be arranged along a single axis, leading to lower values of $s$.
This suggests that care must be taken when using highly biased selections so as not to miss galaxies apherically distributed around the protocluster outskirts.

We see evidence in the evolution of $s$ and $T$ for the emergence of a red sequence.
Between $z = 8.93$ and $z = 3.95$, $\bar{s}$ rises steadily from 0.36 to 0.49, then falls to 0.45 by $z = 2.07$.
There is no dramatic collapse in spatial extent over this period which could explain the fall in $s$ \citep{muldrew_what_2015}; most of the collapse to form current-day clusters occurs at $z < 2$.
Instead, we attribute it to a decrease by a factor of 2 in the number of \SelSFR\ galaxies between $z = 2$ and $3$, with the decrease predominantly toward the center of each protocluster (for which we see evidence in \Fig{compur}): those galaxies that do make the \SelSFR\ cut are distributed irregularly outside the protocluster centre, leading to aspherical distributions.

\subsection{Spherical Profiles}
\label{sec:spherical}

\begin{figure}
	\includegraphics[width=\columnwidth]{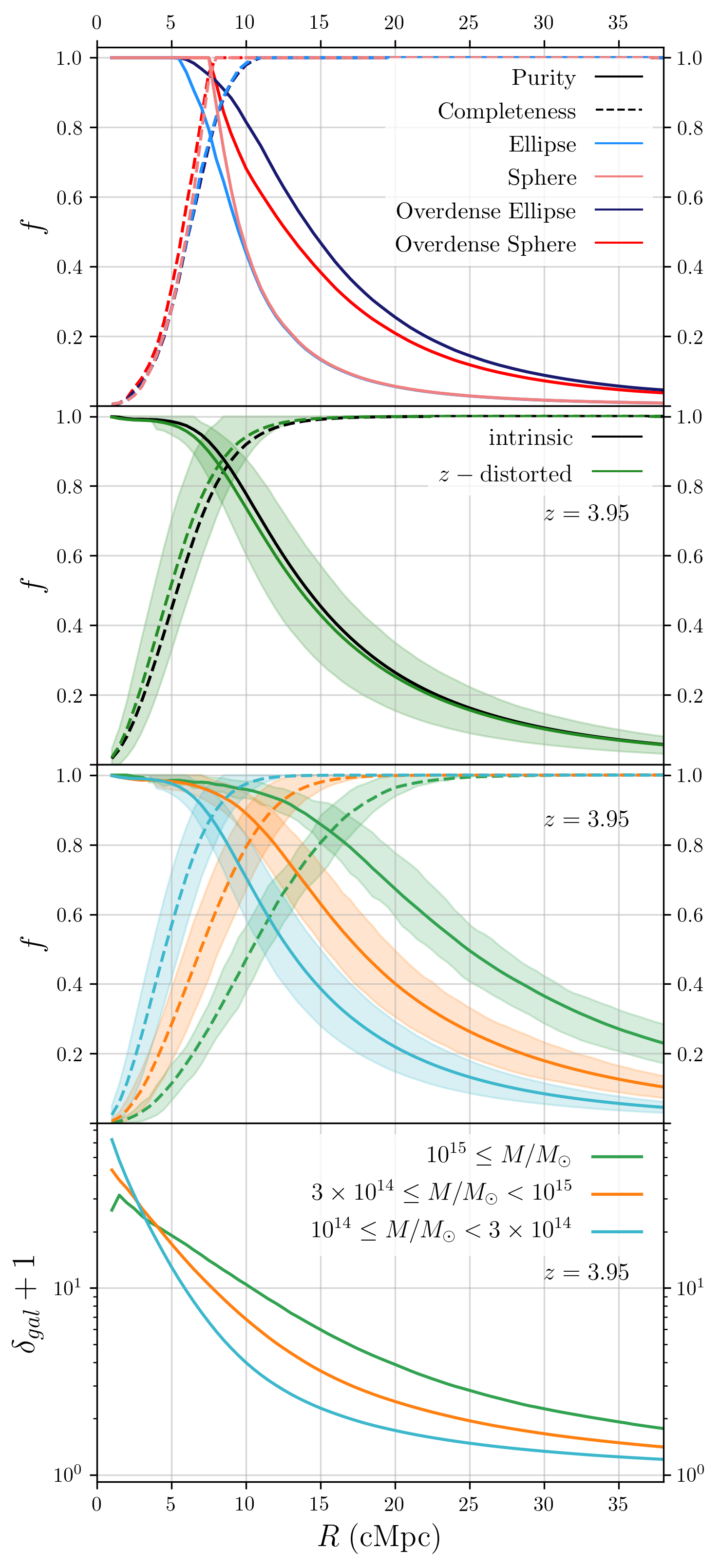}
    \caption{Average spherical profiles of protocluster galaxy properties in comoving coordinates. \textit{Top panel:} Theoretical completeness (dashed) and purity (solid) profiles for a model ellipse and sphere with $\delta_{g} + 1 = 1$ and $\delta_{g} + 1 = 5$. \textit{Second panel:} Mean purity and completeness profiles of the protocluster galaxy population at $z = 3.95$ for the \SelSFR\ selection. Intrinsic (black) and redshift space distorted (green) curves are shown, along with their \nth{16}-\nth{84} percentile range. \textit{Third panel:} The same redshift space distorted profile as in the second panel, split in to three descendant cluster mass bins. \textit{Bottom:} stacked galaxy overdensity profiles (including redshift space distortions), split in to three descendant cluster mass bins.}
    \label{fig:compur_masscomp_dgal}
\end{figure}

\begin{figure}
	\includegraphics[width=\columnwidth]{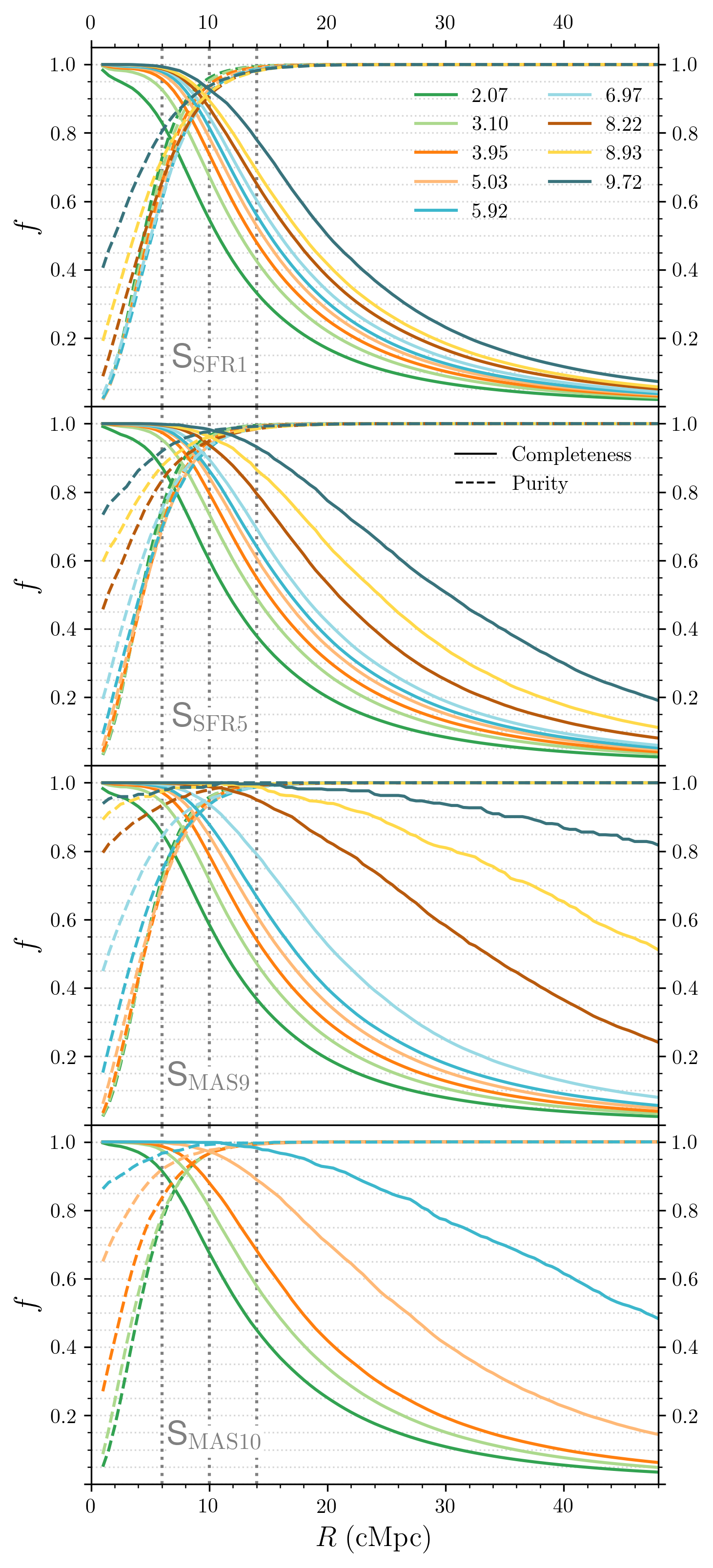}
    \caption{Mean completeness (dashed) and purity (solid) profiles for the protocluster population at a range of redshifts (labelled in the top panel). Panels top to bottom show the \SelSFR, \SelSFRFive, \SelNine\ and \SelTen\ selections, respectively. Vertical dashed lines show the approximate aperture sizes used in \Fig{probabilities_A}. }
    \label{fig:compur}
\end{figure}

Galaxy overdensities are typically measured within cylindrical apertures along the line of sight \citep{franck_candidate_2016}.
\Sec{triaxial} shows that protocluster galaxies tend to be aspherically distributed with a prolate configuration, so such measurements could be biased by the introduction of many field galaxies, or by missing extended protocluster structure not contained within the aperture.
To investigate we measure the properties of protoclusters as a function of radius from their centre (defined as the median coordinates of the selected protocluster galaxies), starting with the completeness and purity profiles of the galaxy population, before moving on to overdensity profiles.

\subsubsection{Protocluster Galaxy Completeness and Purity Profiles}

We begin by looking at the evolution in the completeness and purity of the protocluster galaxy population as a function of radius for a toy model ellipse.
The volume of the ellipse represents the protocluster galaxy distribution, and outside represents the field.
The shape of the model ellipse is based on the mean measured protocluster axis lengths for the \SelSFR\ selection at $z = 3.95$,\footnote{$a = 11.00$, $b = 7.56$ and $c = 5.36 \; (\mathrm{cMpc})$}, and initially assume the galaxy distribution is identical in both protocluster and field.

The purity and completeness as a function of radius can then be derived from the volume ratios, as shown in the top panel of \Fig{compur_masscomp_dgal}.
The model ellipsoid is labelled `Ellipse' and shown in blue, and a spherical model with the same volume is labelled `Sphere' and shown in light pink.
Close to the centre the completeness is low and the purity high, as expected; as the sphere is grown the completeness increases until it encapsulates all of the ellipse, whilst the purity begins to fall as more field volume is included.
The curves cross at high values of both completeness and purity.

The second panel of figure \ref{fig:compur_masscomp_dgal} shows the mean completeness and purity curves for the protocluster galaxy population in \textsc{L-galaxies} at $z = 3.95$.
We define the centre of the protocluster as the median of the protocluster galaxy coordinates, the completeness as the fraction of all protocluster galaxies within the aperture, and the purity as the fraction of galaxies within the aperture that are members of the protocluster.
Both intrinsic (black) and redshift space distorted (green) coordinates are shown.
The \nth{16}-\nth{84} percentile range is shown as a shaded region; the majority of protoclusters exhibit similar profiles, and cross over at high values within a tight range of radii.

The purity and completeness curves both show gradual evolution toward the edge of the protocluster, rather than the sudden change seen in our toy model, and the purity curve drops off much more gradually, which we attibute to our naive assumption of a uniform distribution of galaxies in our toy model -- in reality, protoclusters have a higher overdensity than the surrounding field.
To model this, we increase the number of samples in the ellipse by a factor of 5, simulating a galaxy overdensity of $\delta + 1 \sim 5$.
The completeness and purity curves for this model are shown in the top panel of \Fig{compur_masscomp_dgal} in dark blue, labelled `Overdense Ellipse'; the purity curve falls much more gradually, as seen in the SAM.
Importantly for measurements of galaxy overdensity, the lower number density of galaxies in the field acts to reduce the effect of asphericity on the measured galaxy population, lowering the contamination on the protocluster outskirts and maintaining relatively high purity out to large radii.
It is not unreasonable then, when producing descriptive statistics on the protocluster population, to adopt spherical symmetry above some minimum radius.

The inclusion of redshift space distortions has two effects.
The coherent motions of galaxies as they fall toward the centre of the forming cluster leads to an apparent flattening in their appearance, known as the Kaiser effect \citep{kaiser_clustering_1987}, and we see evidence for it in the steeper completeness curve; galaxies appear closer to the centre, which can be explained if they are, on average, infalling,  \citep{contini_semi-analytic_2016}, and this acts to marginally boost the overdensity measurement.
The purity curve drops at lower radii, which suggests greater apparent contamination from field galaxies; these galaxies are gravitationally disturbed by the forming protocluster, but do not enter the virial radius by $z=0$.
The two curves still cross at high values ($> 80 \%$).

The third panel of \Fig{compur_masscomp_dgal} shows the mean completeness and purity curves for protoclusters at $z=3.95$ split by descendant cluster mass.
There is a positive correlation between cluster size and crossover radius: protoclusters with the most massive descendants trace larger volumes than those that will form lower mass clusters.
In order to capture the majority of the galaxies in the most massive protoclusters a much larger field of view is required.
However, the majority of protoclusters can be captured in their entirety using a much smaller aperture, and even the largest protoclusters contain a significant fraction of their tracer galaxies within a smaller aperture ($>50\%$ at $R = 10 \; \mathrm{cMpc}$).
The crossover values remain high ($>80\%$) for all mass bins.

Figure \ref{fig:compur} shows the mean completeness and purity for each selection criteria with redshift.
For the most stringent selections at the highest redshifts the completeness curves start at non-zero values, since some protoclusters may be represented by only a single galaxy, boosting the mean.
Similarly, the purity curves also remain high, since where galaxies are rare in protoclusters, they also tend to be rare in the field; where they exist, they are highly clustered and located in protoclusters (see \Fig{pc_fractions}).
The purity curve falls at lower radii with decreasing redshift for all selections, caused by the protocluster collapse and central concentration, and the higher relative density of field galaxies with decreasing redshift (see \Fig{pc_fractions}).

The exception to this evolution is seen at low redshift ($z \leqslant 3$) for both \SelSFR\ and \SelSFRFive: the purity falls significantly at much lower $R$, and the completeness curve is also steeper.
\Fig{pc_fractions} shows that the number of \SelSFR\ protocluster galaxies decreases below $z = 3.10$, which can be explained by the emergence of a red sequence; since there are fewer star forming galaxies at the centre of protoclusters relative to the outskirts, the completeness curve rises more rapidly with radius.
We see further evidence for the emergence of a red sequence in the asphericity distribution between $z = 3$ and $2$ (see \Sec{triaxial}).

The crossover between purity and completeness remains high, $\geqslant 80$\,\%, and is relatively insensitive to changes in redshift or selection criteria.
The cross over radii also all fall within a narrow range of values, which suggests a characteristic scale can be chosen,
\begin{equation}
\RC \sim 10 \, \mathrm{cMpc} \;,
\end{equation}
that maximises the completeness and purity regardless of selection criteria or redshift.
This corresponds approximately to an angular scale ($2 \RC$) of 10 arcmin on the sky at $z=2$, falling to 6 arcmin by $z = 9$, not much larger than typical focused searches around biased tracers such as AGN.

\subsubsection{Protocluster Galaxy Overdensity Profiles}
The bottom panel of \Fig{compur_masscomp_dgal} shows the differential stacked overdensity profiles, measured using all galaxies (protocluster+field) within a spherical aperture centred on the protocluster, and split by descendant mass.
We find similar centrally peaked profiles to the surface overdensities measured in \cite{overzier_cdm_2009} \& \cite{chiang_ancient_2013}.
The slope of the overdensity profile at small-intermediate radii is shallower for higher mass protoclusters -- they are less centrally concentrated and more extended -- and for lower mass protoclusters they are more sharply peaked toward the centre.
This may be as a result of our protocluster centre definition: lower mass protoclusters typically have only a single dominant group, so the centre will be defined within this group, leading to a peaked profile at low $R$. Conversely, in larger protoclusters with multiple similarly sized subgroups the median coordinates may lie in an intergroup region, lowering the measured overdensity on small scales.
However, measuring the overdensity centred on a single subgroup will not be representative of the entire protocluster, and may lead to lower purity and completeness at larger radii. We therefore emphasise the need to make descendant mass estimates from overdensity measurements over sufficiently large apertures ($R > 7 \, \mathrm{cMpc}$), which we demonstrate in \Sec{mass}.
The variation in slope of the overdensity profile with mass suggests that measuring overdensity on multiple scales could lead to a more accurate descendant mass estimate, however we found that the improvement in the fit is not substantial.

\subsection{Galaxy Overdensity Statistics}
\label{sec:gods}

Protoclusters have irregular shapes, but this has a small effect on the completeness and purity of their galaxy populations when measured in a sufficiently large aperture.
However, the size and shape of the aperture used to measure the overdensity can have a significant effect on the qualitative value of the overdensity (see the bottom panel of \Fig{compur_masscomp_dgal}, and \cite{shattow_measures_2013}), on which further properties, such as protocluster probabilities and descendant masses, are based.
We propose an improved procedure for deriving overdensities that takes in to account irregular apertures.

\begin{table}
\caption{Candidate region labelling conditions. $C$ is completeness, $P$ purity, and \Clim\ and \Plim\ are limiting values of each that differentiate each classification.}
\label{tab:conditions}
\centering.
\begin{tabular}{|| l | c | p{30mm} ||}
 \hline \hline
 Label & Condition & Description \\
 \hline
 \Proto & $C \geq \,$\Clim$ \; \mathrm{and} \; P \geq \,$\Plim & Protocluster region. \\
 \ProtoF & $C \geq \,$\Clim$ \; \mathrm{and} \; P < \,$\Plim & Region traces the \newline combination of a proto- \newline cluster and field region. \\
 \PProto & $C < \,$\Clim$ \; \mathrm{and} \; P \geq \,$\Plim & Region traces a part \newline of a protocluster. \\
 \Field & $C < \,$\Clim$ \; \mathrm{and} \; P < \,$\Plim & Field region. \\[1ex]
 \hline
\end{tabular}
\end{table}

\subsubsection{Identifying Protoclusters in Galaxy Overdensities}
\label{sec:overdensities}

\begin{figure*}
	\includegraphics[width=\textwidth]{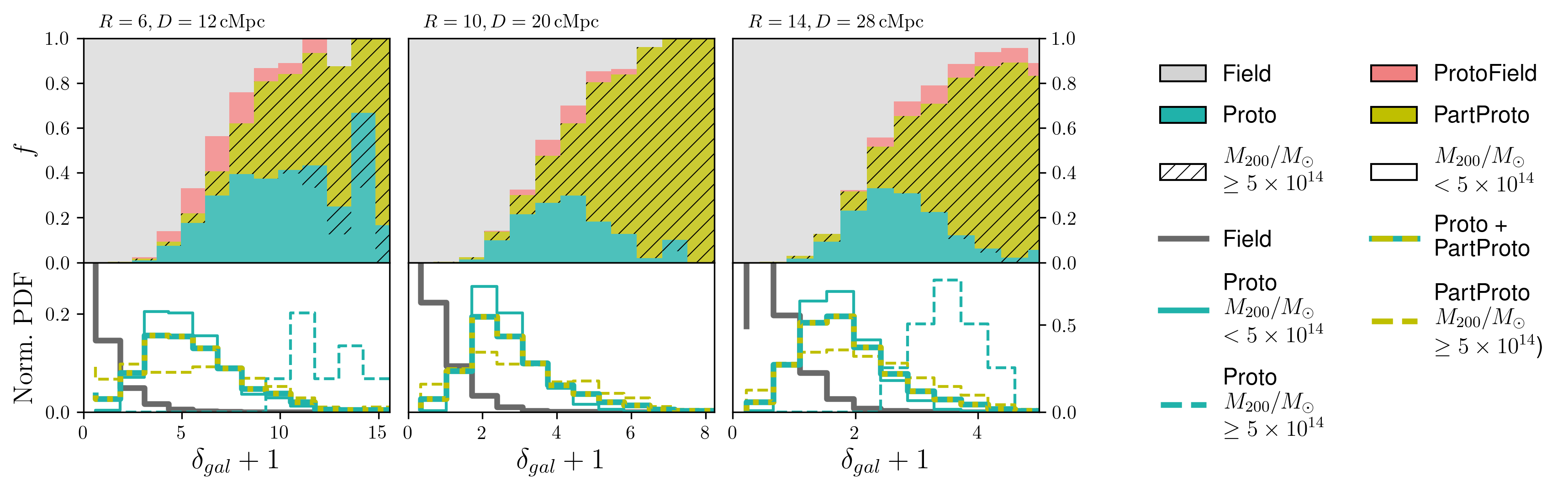}
    \caption{\textit{Top:} Fractional probability distribution of candidate being \Proto, \PProto, \ProtoF\ or \Field\ (\SelSFR, $z=3.95$). Where the distribution is hatched represents those candidates that trace high mass ($M_{200} / M_{\odot} \geq 5 \times 10^{14}$) protoclusters. Each panel shows a different aperture size, labelled at the top. We choose \Clim\ and \Plim\ values equal to the $\mathrm{5^{th}}$ percentile of the completeness and purity of the protocluster population (for this aperture and selection).
    \textit{Bottom:} Normalised probability density distributions for each classification, split into low and high mass descendants.}
    \label{fig:probabilities_A}
\end{figure*}

\begin{figure}
	\includegraphics[width=\columnwidth]{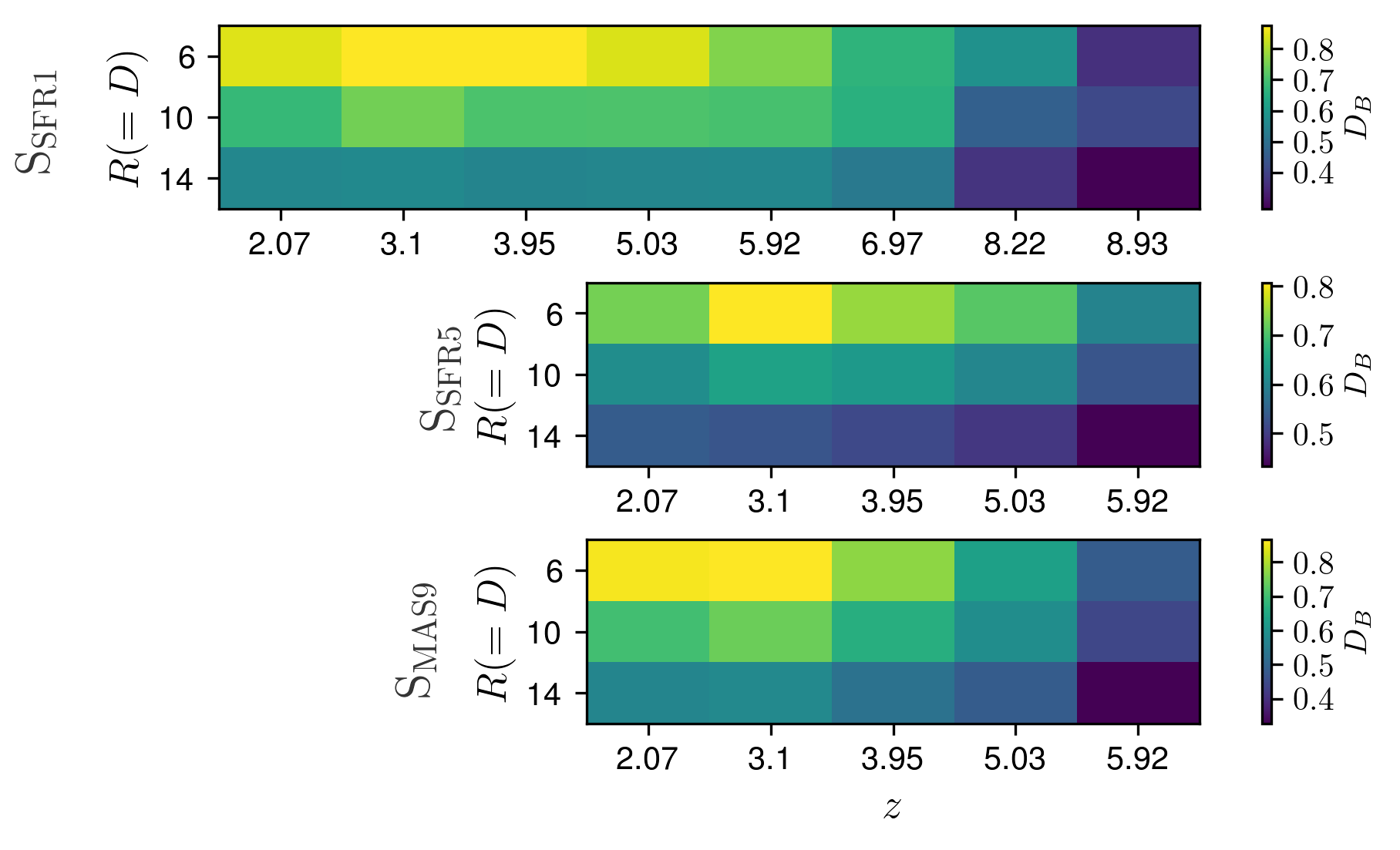}
    \caption{Colour map showing the Bhattacharrya distance ($D_{B}$) between the combined \Proto+\PProto\ and \Field\ distributions for the \SelSFR\, \SelSFRFive\ and \SelNine\ selections, over a range of redshifts ($z$) and aperture sizes ($R = D / 2, \; \mathrm{cMpc}$). The \SelTen\ selection, and some redshifts, are not shown since there are insufficient galaxies to produce a reasonable statistic. $D_{B}$ is maximised at $R = 6$ for all selections at almost all redshifts, and decreases as the selection region is increased in volume.}
    \label{fig:db_map}
\end{figure}

We select $100\,000$ random regions, with surface area, $\pi\,R^{2}$, and depth, $D\equiv\Delta d'$, in the Millennium volume.
We call each of these regions a \textit{candidate}.
For each galaxy in the candidate we find its descendant halo mass.
If no galaxies in the candidate have cluster descendants, the candidate is labelled a field region.
If there are cluster progenitors in the candidate, the completeness, $C$, and purity, $P$, of the galaxy population in this candidate with respect to each descendant cluster is calculated.
Each region can then be classified as \Proto: `protocluster', \ProtoF: `protocluster+field', \PProto: `part of a protocluster', or \Field: `field' according to the conditions detailed in \Tab{conditions}.
In the rare case where there are multiple cluster descendants, the cluster with the highest value of the purity and completeness added in quadrature is chosen.

Importantly, the values of \Clim\ and \Plim\ are chosen based on the $\mathrm{5^{th}}$ percentile of the completeness and purity of the protocluster population \textit{given the chosen selection criteria and aperture}.
This allows a more accurate characterisation of candidate regions that takes into account the actual galaxy membership of protoclusters.
For example, one would not expect to have high purity in a large aperture due to contamination from field galaxies on the outskirts, but would demand high completeness since the majority of a protoclusters galaxies should be captured.
We demonstrate the effect of changing \Clim\ and \Plim\ whilst maintaining a fixed aperture in \App{appendix1}.

Once all candidates are labelled, we can calculate the fractional probability that a measured overdensity represents one of our 4 labels, further split by the mass of the descendant cluster.
\Fig{probabilities_A} shows an example; the upper panel shows the fractional probability distribution, the lower panel the probability density distribution.
The default parameters are $R = D / 2 = 10 \, \mathrm{cMpc}$ and $z = 3.95$, using the \SelSFR\ selection, and we choose \Clim\ and \Plim\ values equal to the $\mathrm{5^{th}}$ percentile of the completeness and purity of the protocluster population with this aperture and selection.
As expected, higher galaxy overdensities are more likely to evolve into clusters, and the highest overdensities are more likely to form more massive protoclusters.
At intermediate to high overdensities, a considerable fraction of candidates trace \PProto\ regions.
All of these \PProto\ candidates trace high mass protoclusters; lower mass protoclusters cannot satisfy \Clim\ whilst simultaneously satisfying \Plim\ as they are not large enough.
At intermediate overdensities there is a small probability that a candidate is probing a \ProtoF\ region, and these are all for smaller, lower mass protoclusters.

The approach is similar to that demonstrated in \cite{chiang_ancient_2013}, though the criterion for classifying a random region as a protocluster is different: they require that the center of the random region lies within half a box length of a protocluster centre, so that the window covers, on average, > 50\% of the protocluster mass.\footnote{private correspondence} Our analysis in \Sec{triaxial} and \Sec{spherical} suggests that the assumption of spherical symmetry is violated, particularly at high-$z$, so this definition may identify regions with significant field galaxy populations.
Despite these differences (including the use of an updated version of \textsc{L-Galaxies} and the Planck cosmology) we achieve consistent results: the protocluster fractions of \SelSFR\ galaxies at $z \sim 4$ match the combined \Proto\ and \PProto\ distribution in the right panel of \Fig{probabilities_A}, with a slight shift in quantitative overdensity to lower values (possibly due to using a slightly larger volume). The probability density distribution for low mass protoclusters appears to show less distinction from the field distribution as seen in Figure 6 in \cite{chiang_ancient_2013}, which may be attributed to the updated protocluster definition, or to the change in cosmology.\footnote{The Planck cosmology used in \cite{henriques_galaxy_2015} leads to an increased dark matter particle mass, an increased box size, and the $z = 0.12$ output of the original WMAP1 simulation becomes the new $z=0$; the latter two effects would lead to a diluted quantitative overdensity measurement}
Whilst consistent, we note that our approach explicitly distinguishes protoclusters identified partially or in whole, and can handle irregularly shaped apertures.

The probability density distributions at the bottom of each panel can be used to evaluate the separation in overdensity space of protocluster and field regions.
We determine the Bhattacharyya distance \citep{bhattacharyya_measure_1946}, a measure of the dissimilarity between two probability distributions, defined as
\begin{equation}
D_{B} = - \ln BC,\ \ \ \mathrm{where}\ BC(p,q) = \sum_{\delta \in \Gamma} \sqrt{p(\delta)q(\delta)}
\end{equation}
and $p$ and $q$ are the probability distributions over the galaxy overdensity domain $\Gamma$.
$D_{B}$, calculated between the \Field\ and combined \Proto\ and \PProto\ distributions for a range of redshifts, aperture sizes and selections, is shown in \Fig{db_map}.
At lower redshifts the distinction is greatest on small scales ($R = 6 \, \mathrm{cMpc}$) for all selections, though the distinction on the characteristic scale ($R = \RC = 10 \, \mathrm{cMpc}$) is still relatively high compared to larger scales.
At higher redshifts the distinction is greatest at $\RC$.
This seems to suggest that, in order to best separate protoclusters from the field, one should use a smaller aperture at lower redshifts and a slightly larger one at higher redshifts.
However, the overdensity profiles shown in \Fig{compur_masscomp_dgal} show that a larger aperture allows the greatest discrimination of protocluster descendant mass, and in \Sec{agn} we find that, in searches surrounding AGN, $D_B$ is maximised at $\RC$ due to the non-central location of the AGN within the protocluster.
We therefore still recommend making overdensity measurement on a scale of $\RC$ for all redshifts and selections.

\subsubsection{Protocluster Mass from Galaxy Overdensity}
\label{sec:mass}

\begin{figure}
	\includegraphics[width=\columnwidth]{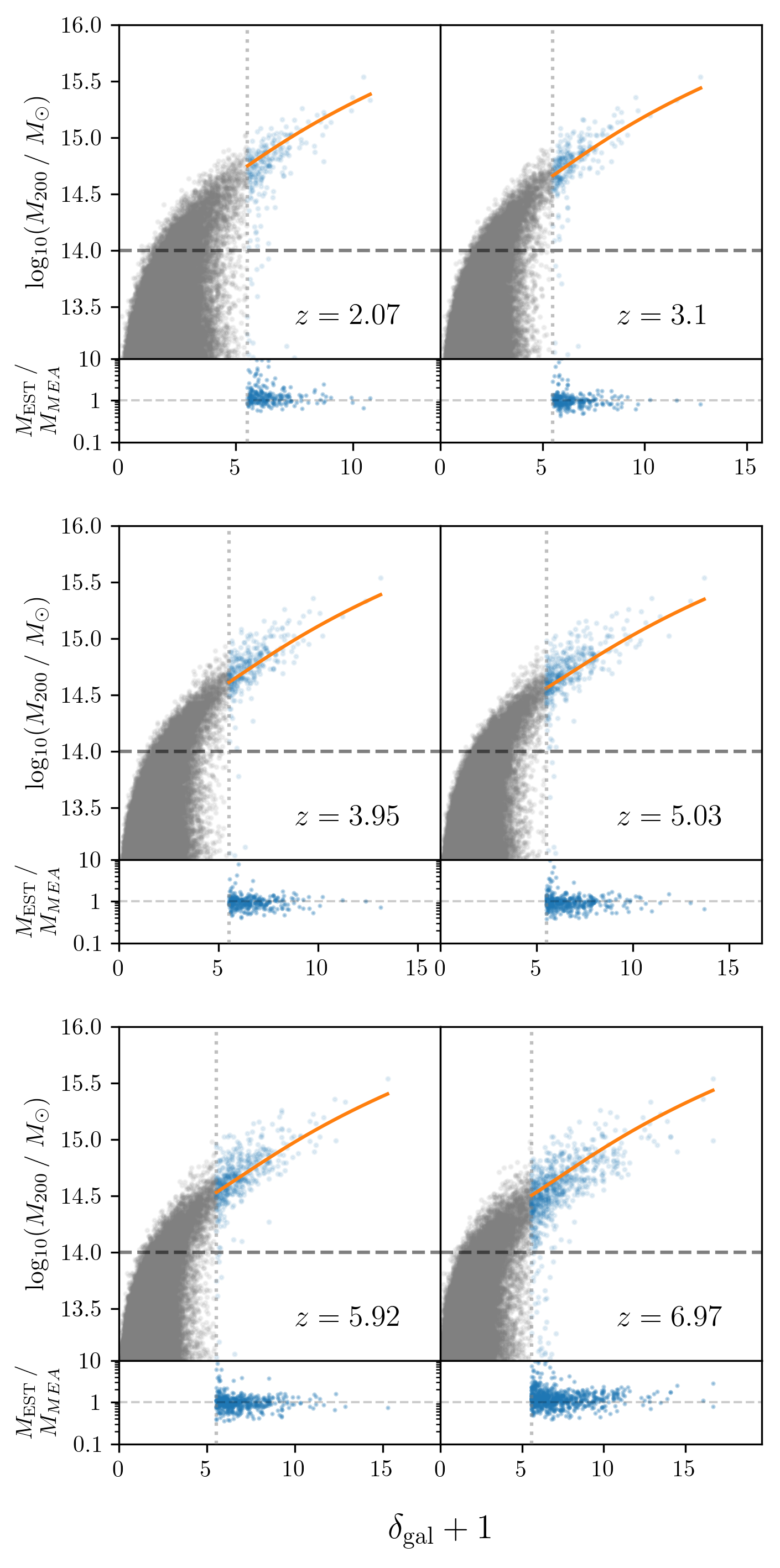}
    \caption{\textit{Top panels:} Galaxy overdensity (\SelSFR) against descendant halo mass for all halos with $\mathrm{log_{10}}(M_{200} \,/\, M_{\odot}) \,>\, 13$. The fit at each redshift is shown in orange. Those objects used in the fit are shown in blue, those below the overdensity threshold in grey. Our cluster mass definition ($\mathrm{log_{10}}(M_{200} \,/\, M_{\odot}) > 14$) is delimited by the horizontal dashed black line. \textit{Bottom panels:} Ratio of the estimated and measured masses.}
    \label{fig:mass_fit}
\end{figure}

\begin{table}
\caption{Protocluster mass estimate fit parameters for Equation \ref{eq:1}, for the \SelSFR, \SelSFRFive\ and \SelNine\ selections, with error estimates.}
\label{tab:fit_params}
\centering
\begin{tabular}{|| c | c c c c c||}
 \hline
 Selection & a & b & c & C & $R^{2}$\\ [0.5ex]
 \hline\hline
 \SelSFR\ &  0.146 & -1.077 & 2.628 & 1.752 & 0.547 \\
 \SelSFRFive\ & 0.658 & -1.317 & 1.859 & 0.117 & 0.549 \\
 \SelNine\ & 2.883 & -1.681 & 1.452 & -0.235 & 0.507 \\ [1ex]
 \hline
\end{tabular}
\end{table}

We now explore the relationship between high redshift overdensity and descendant cluster mass by fitting an empirical relation between the two.
We fit to all halos at $z=0$ with masses $M_{200} / M_{\odot} > 10^{13}$ in order to fully assess the spread in descendant masses for a given overdensity, calculating the overdensity measured in a single cylindrical aperture with radius and depth equal to the characteristic scale, $\RC = 10 \; \mathrm{cMpc}$; on smaller scales descendant mass cannot easily be distinguished through galaxy overdensity (see \Fig{compur_masscomp_dgal}, bottom panel).

The relationship between overdensity and descendant mass is parameterised as follows:
\begin{equation} \label{eq:1}
M_{200} / (10^{14} M_{\odot}) = a \, (1+z)^b \, (1+\delta)^c \,+\, C \;.
\end{equation}
where $M_{200}$ is the descendant mass, and $\delta$ is the measured galaxy overdensity.
We fit the \SelSFR, \SelSFRFive\ and \SelNine\ distributions between $z = 2-7$ using the \texttt{curve\_fit} least squares minimisation method provided by \textsc{scipy} \citep{scipy}.
The fit is illustrated in \Fig{mass_fit} for the \SelSFR\ selection, with residuals shown at the bottom of each panel.
We ignore both the \SelTen\ selection and $z > 7$ due to a lack of galaxies.
A striking feature of \Fig{mass_fit} is the spread in descendant halo masses for $\delta_{gal} < 4.5$.
We cannot make any meaningful descendant mass prediction below this overdensity limit, so we limit our fit to above this range; whilst there is a chance that such regions do trace protoclusters, the vast majority of them do not.
The exact choice of threshold overdensity depends on many factors that affect the overdensity distribution (aperture size, selection, etc.).
For this aperture, the distribution conveniently turns over at descendant masses of $\sim 10^{14} \, M_{\odot}$, which makes distinguishing high mass protoclusters by overdensity somewhat easier; lower mass protoclusters are harder to distinguish from the field.

A non-linear relationship provides a marginally better fit for the very highest descendant masses.
In \Sec{spherical} we noted that the shape of protocluster overdensity profiles was dependent on their descendant mass, but including overdensity measured on two scales leads to no appreciable improvement in the fit, which we attribute to the scatter in overdensity profiles.

\cite{chiang_ancient_2013} derive a similar relation between overdensity and descendant mass, ignoring redshift space distortions, but taking in to account the aperture size, whilst the coefficients of our empirical model must be rederived for differing apertures.
We note that they only apply it to overdensities surrounding protoclusters, which underestimates the scatter in descendant halo mass at intermediate overdensities (see \Fig{mass_fit}), and in their Figure 12 showing the residuals they ignore objects with descendant masses below the protocluster mass threshold.

\subsection{AGN as Protocluster Tracers}
\label{sec:agn}

Both quasars and HzRGs are expected to act as tracers of protocluster regions. In order to test this assumption we select a sample of quasars and HzRGs whose number densities match observations at high-$z$ (\Sec{agn_sel}), find their surrounding galaxy overdensities (\Sec{agn_od}) and investigate their coincidence with protoclusters (\Sec{agn_frac}).

\subsubsection{AGN selection}
\label{sec:agn_sel}

\begin{figure}
	\includegraphics[width=\columnwidth]{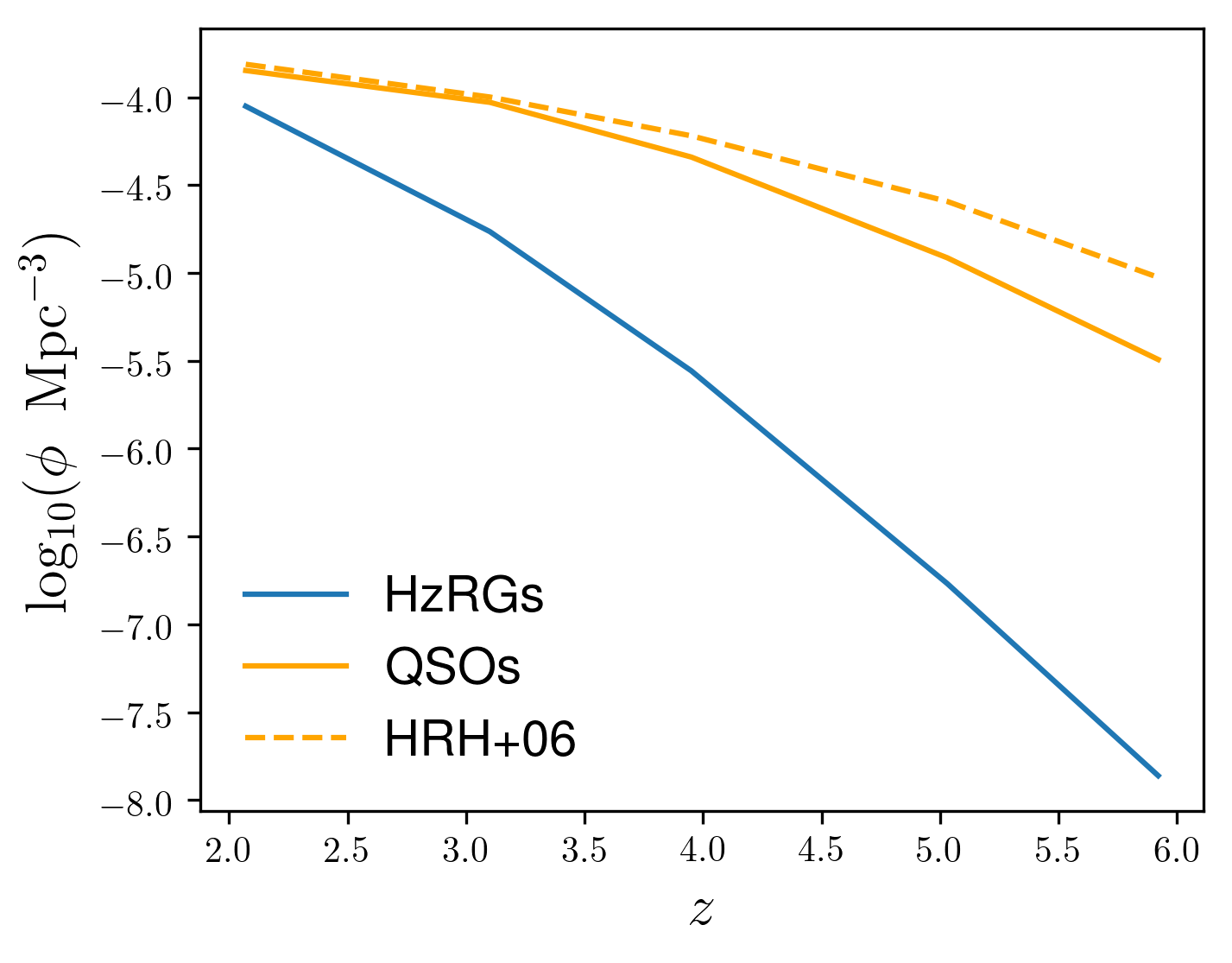}
   \caption{Number density evolution of HzRGs (blue) and quasars (solid orange) subject to the accretion cuts stated in \Sec{agn}. The quasar mode accretion cut was selected in order to match the number density evolution as measured by \protect\cite{hopkins_observational_2007} (dotted orange).}
   \label{fig:AGN_number_densities}
\end{figure}

We choose our quasar mode accretion cut in order to match the integrated number densities from \cite{hopkins_observational_2007} between $z = 2 - 5$ (assuming a lower luminosity limit of $10^{44} \; L_{\mathrm{bol}} \,/\, \mathrm{erg \; s^{-1}}$):
\begin{equation}
\dot{M}_{\mathrm{\bullet}}(\mathrm{quasar}) / (M_{\odot} \, \mathrm{yr^{-1}}) >  0.0036 \;.
\end{equation}
This gives a reasonably good fit to the normalisation and redshift evolution (see \Fig{AGN_number_densities}).
The accretion rate can be translated into a bolometric luminosity through the following prescription,
\begin{equation}
L_{\mathrm{bol}} = \epsilon \dot{M}_{\bullet} c^{2}
\end{equation}
where $\dot{M}_{\bullet}$ is the accretion rate and $\epsilon = 0.1$.
For the quasar accretion mode this gives a lower limit of $L_{\mathrm{bol}} > 2 \times 10^{43} \, \mathrm{ergs \; s^{-1}}$, somewhat lower than typical intermediate-luminosity quasars, which suggests an underprediction of the black hole accretion rate at high-$z$.

\Fig{AGN_number_densities} shows a similar decline in number density of HzRGs in the model from $z \sim 2$.
There is still significant uncertainty about the position and luminosity dependence of a high redshift cutoff in observations \citep{jarvis_redshift_2001, venemans_protoclusters_2007, rigby_luminosity-dependent_2011}; we therefore choose a radio mode accretion threshold in order to approximately match the number densities measured by \cite{dunlop_redshift_1990} for the most powerful radio galaxies over the redshift range $z = 2 - 5$:
\begin{equation}
\dot{M}_{\mathrm{\bullet}}(\mathrm{radio}) / (M_{\odot} \, \mathrm{yr^{-1}}) > 0.001 \;.
\end{equation}
We also adopt more conservative accretion cuts in order to test any dependence on the chosen cut-off (see \Sec{agn_od}).

\begin{figure}
	\includegraphics[width=\columnwidth]{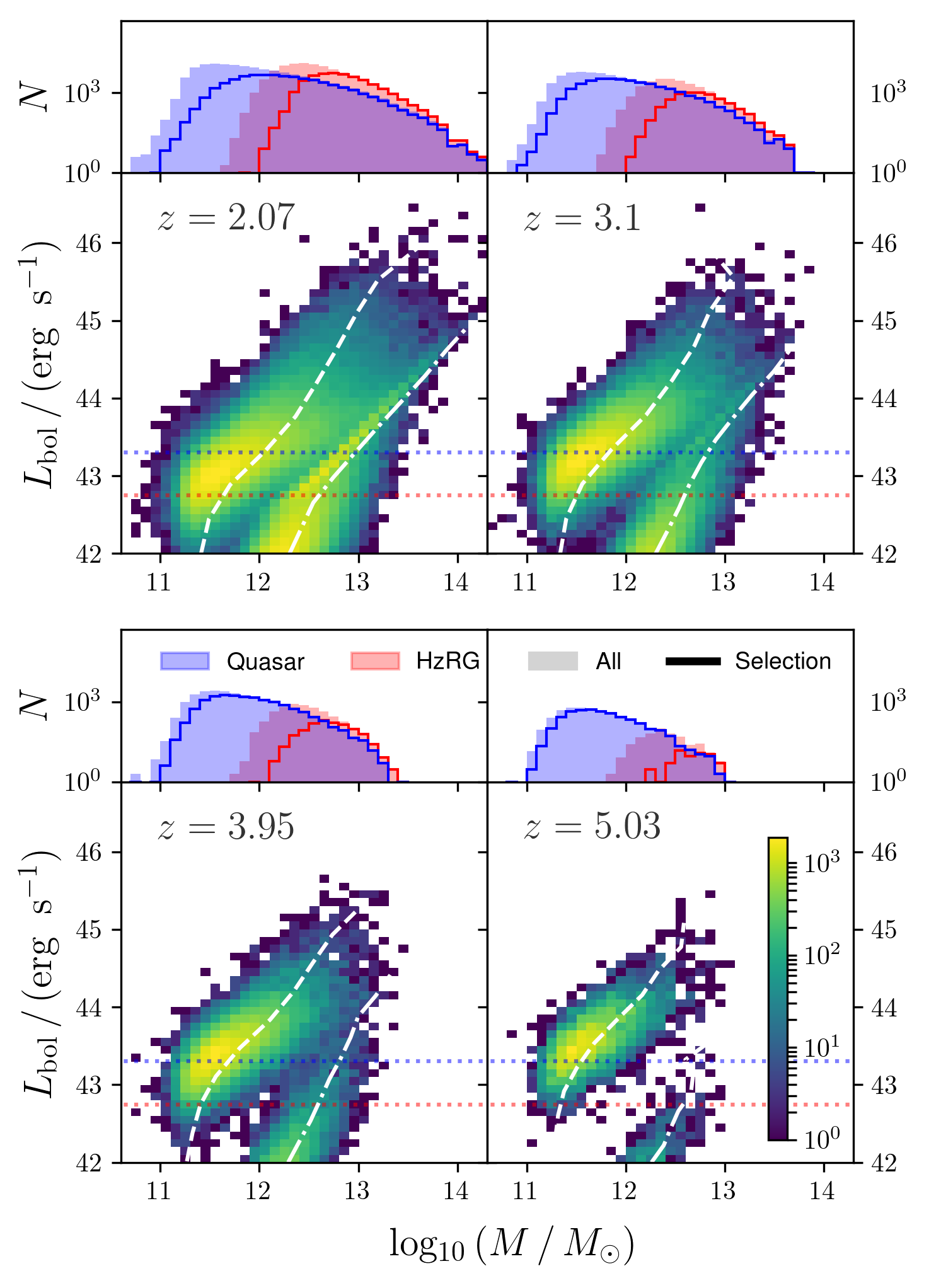}
   \caption{Distribution of AGN luminosity against host halo mass, for a range of redshifts. \textit{Bottom panels:} 2D distribution of bolometric luminosity for the combined radio \& quasar accretion modes against host halo mass. White dashed and dash-dot lines show the independent median of the relationship for the \textit{quasar} and \textit{radio} accretion modes, respectively. Horizontal red and blue dashed lines delimit the accretion cuts stated in \Sec{agn}. \textit{Top panels:} Marginal distribution of host halo masses for the whole AGN population as filled histograms, and as step histograms for the accretion cuts stated in \Sec{agn}.}
   \label{fig:AGN_luminosities}
\end{figure}

Each panel of \Fig{AGN_luminosities} shows the distribution of black hole accretion rates as a function of host halo mass, for a range of redshifts, along with the marginal distribution of halo masses for the total AGN population and our selection.
Each accretion mode is distinct: HzRG tend to populate higher mass halos, with a median mass $\mathrm{log_{10}}(M \,/\,  M_{\odot}) \sim 12.5$, as expected since it is only the most massive halos that have a sufficient reservoir of hot gas to power this accretion mode.
Quasars populate a much wider range of halo masses with a lower median mass of $\mathrm{log_{10}}(M \,/\,  M_{\odot}) \sim 11.5$ at all redshifts considered.
The quasar mode accretion rate is proportional to the product of the ratio of the masses of the merging galaxies and their combined cold gas mass, $\dot{M}_{\mathrm{\bullet}}(\mathrm{quasar}) \propto M_{\mathrm{sat}} \,/\, M_{\mathrm{cen}} \times M_{\mathrm{cold}}$.
Whilst major mergers of high mass halos are rare, high quasar mode accretion rates can still be achieved in massive halos through minor mergers where the primary halo has a large gas reservoir.

\subsubsection{Galaxy Overdensities Surrounding AGN}
\label{sec:agn_od}

\begin{figure*}
	\includegraphics[width=\textwidth]{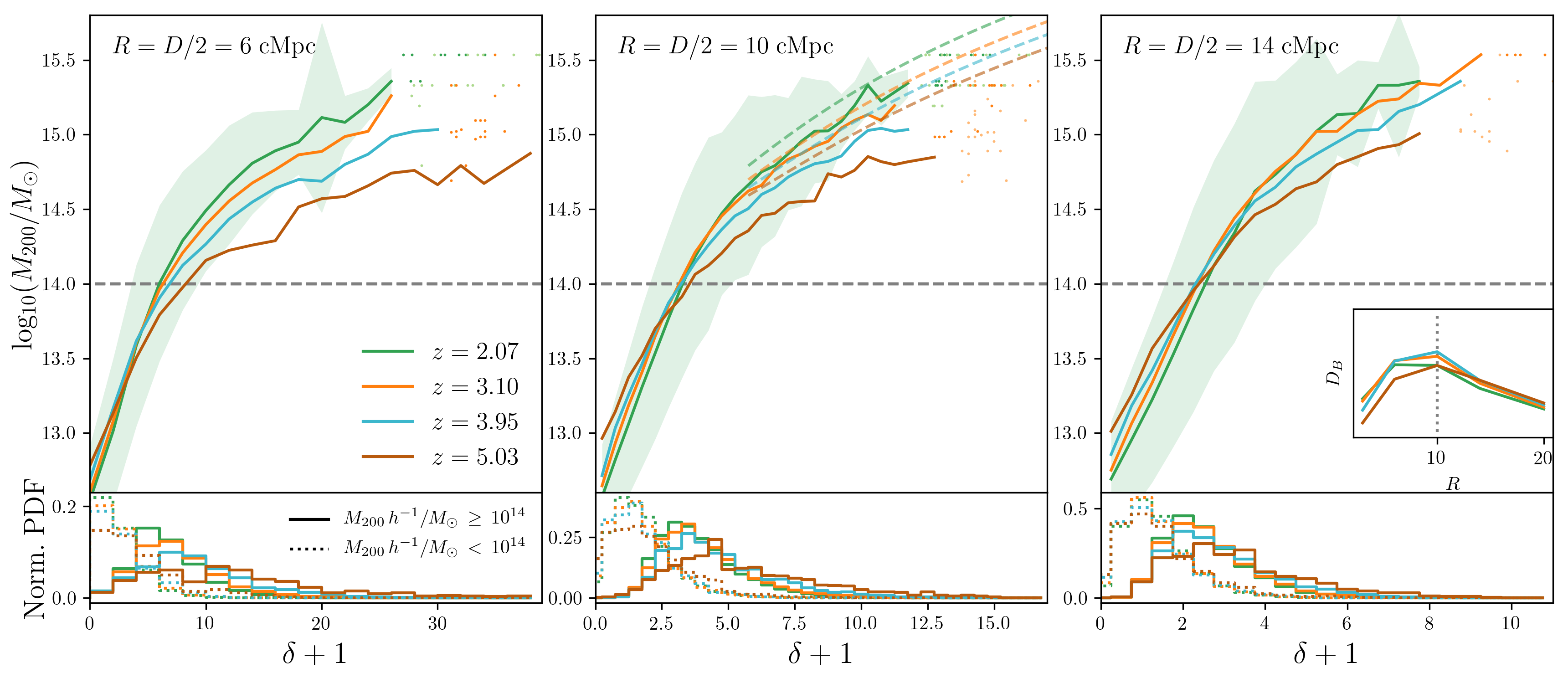}
    \caption{\textit{Top:} Galaxy overdensity (\SelNine) in the vicinity of quasars (selected according to the criteria in \Sec{agn}) against descendant halo mass. Solid lines show the binned mean, and the shaded region shows the \nth{16}-\nth{84} percentile range for the $z=2$ selection. Where there are less than 20 quasars in a bin, individual objects are plotted. The fit from \Sec{mass} is shown as the dashed line in the central panel. \textit{Bottom:} Probability density functions (PDF) for those quasars that evolve into clusters, and those that do not. \textit{Inset:} Bhattacharrya distance, $D_{B}$, between the PDF for quasars that evolve in to clusters and those that do not, as a function of aperture size. The peak indicates the aperture size at which AGN embedded in protoclusters are best discriminated from the field.}
    \label{fig:quasar_select}
\end{figure*}

\begin{figure*}
	\includegraphics[width=\textwidth]{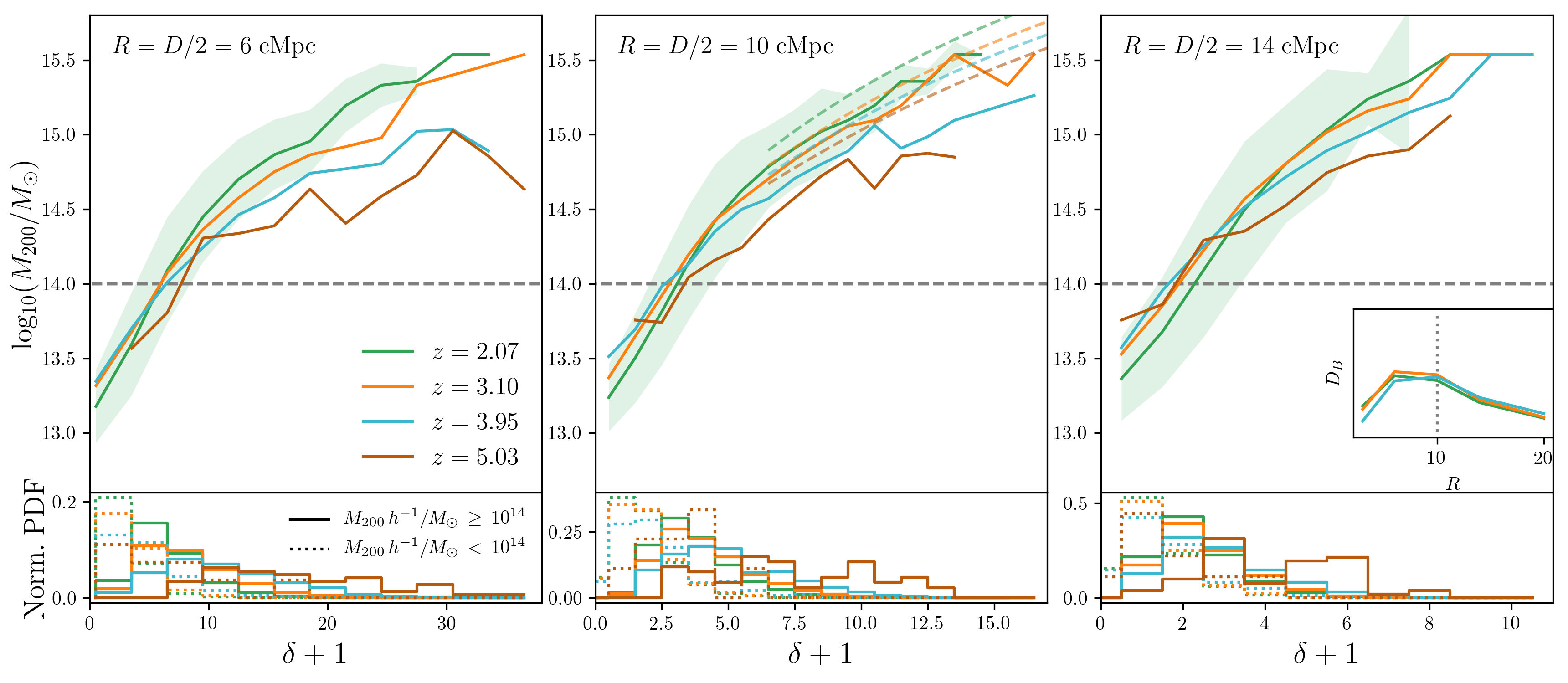}
    \caption{As for \Fig{quasar_select}, but for the HzRG selection.}
    \label{fig:radio_select}
\end{figure*}

Given our AGN selection criteria from \Sec{agn_sel}, figures \ref{fig:quasar_select} and \ref{fig:radio_select} show the galaxy overdensity (\SelNine) in the vicinity of each quasar and HzRG (respectively) against its descendant halo mass for a range of redshifts and aperture sizes.\footnote{for brevity we use regular apertures, $R = D/2$.}
Each coloured line shows the binned mean for all AGN at each redshift, and $\mathrm{16^{th} - 84^{th}}$ percentiles are shaded for the $z=2$ selection.
These figures can be used to read off the estimated descendant halo mass of an AGN given its surrounding galaxy overdensity.

The bottom of each panel shows the normalised probability density distribution for those AGN that end up in clusters and those that do not, in solid and dotted lines respectively, which can be used to calculate the Bhattacharrya distance (introduced in \Sec{overdensities}) to evaluate their level of separation in overdensity space.
$D_{B}$ is shown as a function of $R$ in the inset figure in the third panel of each figure; it peaks between $5 \,-\, 10 \; \mathrm{cMpc}$ for both selection, but slightly higher for quasars.
This is also higher than that seen for random regions of the same size in \Sec{overdensities}; this can be explained by the non-central location of AGN within protoclusters.
For protoclusters containing quasars, the median distance of the quasar from the centre is $\sim 5.05 \, \mathrm{cMpc}$ at $z =3.95$; apertures of size $\sim 10 \; \mathrm{cMpc}$ capture the greatest proportion of the overdense protocluster whilst minimising field contamination, boosting the overdensity associated with that AGN, whereas smaller apertures sample the low overdensity tail.
For HzRGs we see a similar trend with radius, but $D_{B}$ peaks at lower radii, which can be attributed to the fact that the median distance of HzRGs from the centre of their host protocluster is lower ($3.04 \; \mathrm{cMpc}$ at $z =3.95$).
\cite{hatch_why_2014} find that radio loud AGN appear to reside in average overdensities on scales of 0.5\,Mpc, but overdense environments on larger scales, in agreement with this interpretation.

The location of each AGN type within protoclusters can be explained by their differing treatment in the model.
HzRGs preferentially appear in higher mass halos; during cluster assembly a dominant subhalo, with mass $M / M_{\odot} \sim 10^{12}$ emerges at intermediate redshifts \citep{chiang_ancient_2013}, typical of HzRG hosting halos, and will either already be at the center of the protocluster region or will migrate towards it.
In contrast, high luminosity quasars can be triggered by both major and minor merger activity; whilst there will be many minor mergers with massive halos in the dominant subhalo, there will also be a large number of major mergers between intermediate mass halos elsewhere in the protocluster, so that the average quasar location is further from the protocluster centre.

The mass predictions from \Sec{mass} are shown as dashed lines in the centre panel.
Puzzlingly, the predicted descendant mass for a given overdensity is lower for AGN than protoclusters: one would expect, for a given protocluster, the centrally measured overdensity to be larger than from the non-central AGN.
We attribute this to a selection effect; not all protoclusters contain AGN at these redshifts, so the selection does not necessarily have the same descendant mass distribution.

\subsubsection{The Coincidence of AGN \& Protoclusters}
\label{sec:agn_frac}

\begin{figure}
	\includegraphics[width=\columnwidth]{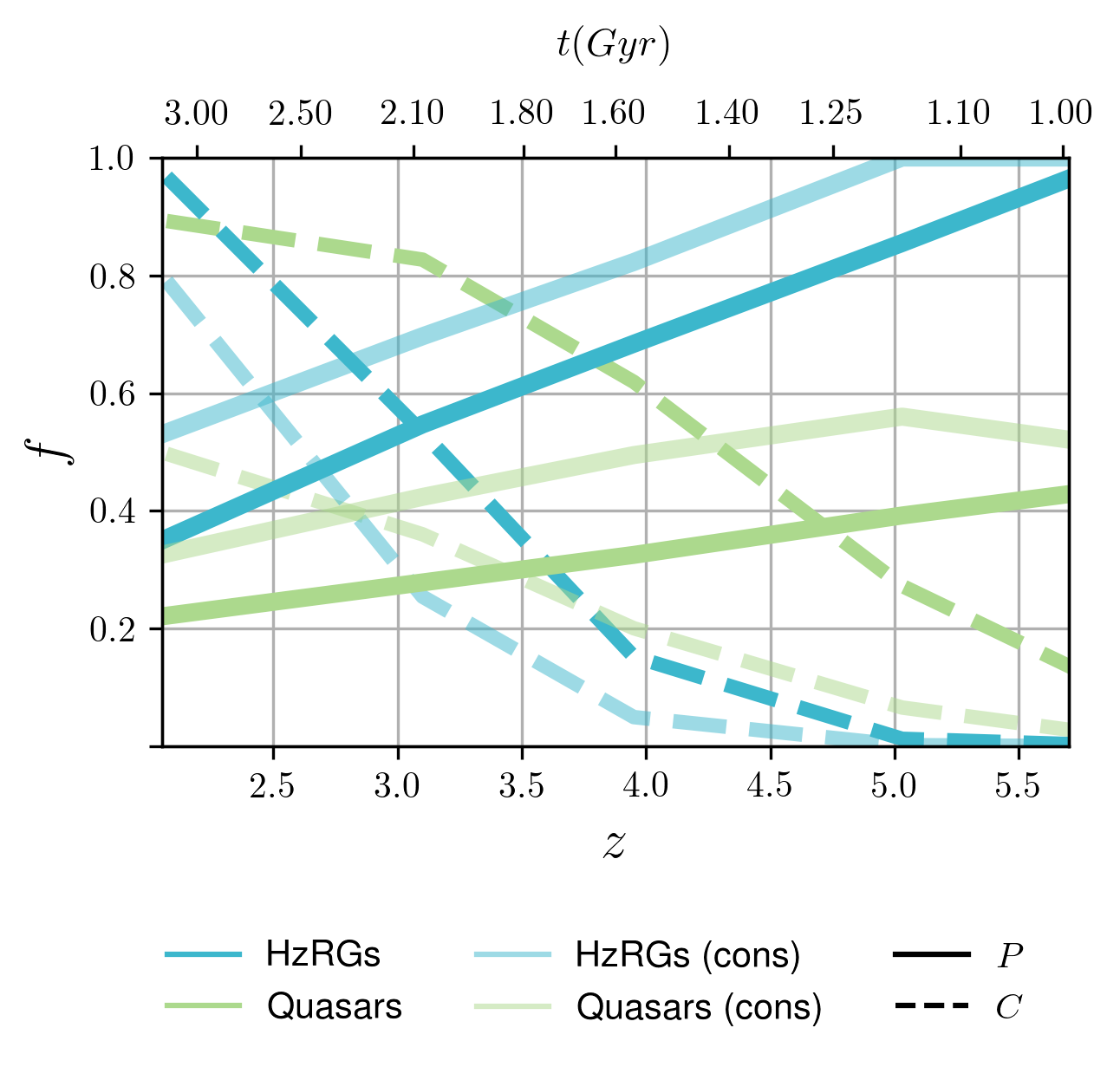}
    \caption{The completeness (dashed), and purity (solid) of AGN as protocluster tracers, for both HzRGs (blue) and quasars (green), and for both accretion thresholds (see \Sec{agn_sel} and \Sec{agn_frac}).}
    \label{fig:agn_stats}
\end{figure}

Figure \ref{fig:agn_stats} shows the \textit{completeness} and \textit{purity} of AGN as biased tracers of protoclusters, where \textit{completeness} in this context refers to the fraction of all protoclusters traced by AGN, and \textit{purity} to the ratio of protoclusters to field regions traced.
In order to assess the effect of our accretion cut choice, we also show the following more conservative accretion cuts:
\begin{align}
\dot{M}_{\mathrm{\bullet}}(\mathrm{radio}) / (M_{\odot} \, \mathrm{yr^{-1}}) &> 0.004 \\
\dot{M}_{\mathrm{\bullet}}(\mathrm{quasar}) / (M_{\odot} \, \mathrm{yr^{-1}})  &> 0.018 \;.
\end{align}
For both selections, at low redshifts the completeness tends to be high and purity low, whilst at high redshift the completeness is low and purity high.
Only at a few intermediate redshifts are the completeness and purity simultaneously high, and this cross over is highly dependent on the adopted accretion threshold.

These trends can be explained by the average host halo mass of quasars and HzRGs.
The massive halos that host HzRGs are the very peaks of the matter distribution at $z > 3.5$, tracing those regions that are most likely to form clusters (see \Sec{mass}), hence the high purity of the selection.
By $z \sim 2$ halos of mass $\mathrm{log_{10}}(M \,/\,  M_{\odot}) \sim 12.5$ are more numerous and do not necessarily coincide with protocluster regions, so the purity decreases, but the completeness rises sharply.
We see no clear evidence for environmental triggering of HzRGs, as suggested by \cite{hatch_why_2014}; instead, HzRGs occur within a narrow range of host halo masses, coincident with forming protocluster cores or groups \citep{chiang_galaxy_2017}.

Similarly, at $z > 5$ the majority of high stellar mass (\SelTen) galaxies reside in protoclusters (see \Fig{pc_fractions}), so major mergers between such galaxies, triggers of quasar mode accretion, will predominantly occur in protocluster environments, hence the high purity of quasar tracers.
This is true of both accretion cuts; the most luminous quasars at $z \sim 6$ do indeed reside in protoclusters, but there are far too few of them to trace an appreciable number of protoclusters.
At later times there is also a population of massive galaxies in the field that may merge, reducing the purity.
There are also less frequent mergers between massive galaxies in protoclusters once a dominant subhalo has formed at the core, which could be responsible for the plateau in completeness at low redshifts.

\cite{orsi_environments_2016} find similar trends in their model; they observe that half of all HzRGs at $z = 2.2$ have cluster descendants, whereas in our model the fraction is approximately between a third and a half, depending on the accretion threshold.
They also find 19\% of quasars have cluster descendants, similar to our value of $\sim 21\%$ for the standard accretion threshold, but slightly lower than the conservative cut.
Observationally, \cite{venemans_protoclusters_2007} find that 75\% of powerful HzRGs in the redshift range $2 \leqslant z \leqslant 5$ reside in protoclusters, which agrees approximately with the mean AGN fraction in this range for both accretion thresholds.
They use a $\sim 3 \times 3$ Mpc aperture, much smaller than $\RC$; the analysis in \Sec{agn_od} suggests that measuring overdensity around HzRGs on this scale will be biased lower, which makes their high measured protocluster fraction somewhat surprising, however they do adopt a more lenient protocluster definition (a factor of 2-5 overdense compared to the field; \Fig{radio_select} suggests an overdensity $>8$ is required) and observe very powerful HzRGs which may be biased toward high mass protoclusters with higher probabilities.
The Clusters Around Radio-Loud AGN (CARLA) survey \citep{wylezalek_galaxy_2013} found 66\% of HzRGs reside in overdense regions at $z \sim 2.4$ \citep{hatch_why_2014}, approximately equal to the conservative accretion threshold, and \cite{wylezalek_galaxy_2013} find 55\% of HzRGs are overdense by $2\sigma$, and 10\% by $\geqslant 5 \sigma$ (for $1.2 < z < 3.2$), which, if we assume that the lower overdensity limit corresponds to true protoclusters, matches our conservative accretion threshold, and the results of \cite{orsi_environments_2016}.

How effective are AGN as biased tracers of protoclusters?
Our model suggests that it depends strongly on redshift.
At high redshift, HzRGs act as reliable tracers of protocluster regions but will not reveal the presence of all protoclusters, whereas quasars reside in a more diverse range of environments.
At lower redshifts almost all protoclusters have at least one AGN, but most AGN do not reside in protoclusters.
At extremely high redshifts, \Fig{pc_fractions} suggests that using masssive galaxies as tracers will lead to the identification of a much more complete sample of protoclusters compared to using AGN, though it should be noted that such galaxies will typically exhibit observable AGN activity too.
We leave the investigation of whether AGN-hosting protoclusters are a distinguishable population for future work.

\begin{table*}
  \caption{Estimated protocluster probabilities for candidates from the literature. All candidate estimates use the $S_{\mathrm{SFR}}$ selection, and combine the Proto and PartProto selections in the protocluster definition. Descendant mass estimates are omitted where protocluster probabilities are low.
  \newline \textbf{Notes:} (a) Redshift. (b) Full width redshift uncertainty. (c) Aperture length corresponding to redshift uncertainty. (d) Observation window area in square arc minutes. (e) Aperture radius giving equal area to the observation window. (f) Measured galaxy overdensity within the specified aperture. (g,h) Mean completeness and purity for each selection, and $5^{\mathrm{th}} - 95^{\mathrm{th}}$ percentile range. We use the lower percentile as our value for $C_{\mathrm{lim}}$ and $P_{\mathrm{lim}}$. (i) Derived protocluster probability. (j) Descendant masses estimated using our fitting procedure.
  \newline \textbf{References:}  (1) \protect\cite{venemans_protoclusters_2007} (2) \protect\cite{steidel_spectroscopic_2005}
  (3) \protect\cite{hatch_halpha_2011} (4) \protect\cite{tanaka_discovery_2011} (5) \protect\cite{venemans_properties_2005} (6) \protect\cite{matsuda_large_2005} (7) \protect\cite{steidel_ly_2000} (8) \protect\cite{yamada_panoramic_2012} (9) \protect\cite{venemans_most_2002} (10) \protect\cite{venemans_discovery_2004} (11) \protect\cite{ouchi_discovery_2005} (12) \protect\cite{toshikawa_discovery_2012}
  }
  \label{tab:obs_chiang}
  \centering.
  \begin{tabular}{|| l | c | c | c | c | c | c | c | c | c | c | c ||}
   \hline \hline
   Name & $z^{\;\mathrm{a}}$ & $\Delta \, z^{\;\mathrm{b}}$ & $D^{\;\mathrm{c}}$ & Window$^{\;\mathrm{d}}$ & $R^{\;\mathrm{e}}$ & $\delta_{g}^{\;\mathrm{f}}$ & $C_{\mathrm{lim}}^{\;\mathrm{g}}$ & $P_{\mathrm{lim}}^{\;\mathrm{h}}$ & $P_{C}(\mathrm{S_{SFR1}})^{\;\mathrm{i}}$ & $\mathrm{log_{10}} (M_{z=0} / M_{\odot})^{\;\mathrm{j}}$ \\
    & & & $\mathrm{cMpc}$ & $\mathrm{arcmin^{2}}$ & $\mathrm{cMpc}$ & & & & & & \\
   \hline

   PKS 1138-262$^{1}$ & 2.16 & 0.053 & 72.6 & 49 & 6.36 & $3^{+2}_{-2}$ & $0.92^{1.0}_{0.60}$ & $0.28^{0.50}_{0.15}$ & 50\% & 14.530 \\[5pt]

   HS1700-FLD$^{2}$ & 2.3 & 0.03 & 38.7 & 64 & 7.52 & $6.9^{+2.1}_{-2.1}$ & $0.98^{1.0}_{0.72}$ & $0.34^{0.59}_{0.18}$ & 100\% & 15.089 \\[5pt]

   4C 10.48$^{3}$ & 2.35 & 0.046 & 58.0 & 6.25 & 2.37 & $11^{+2}_{-2}$ & $0.3^{0.6}_{0.08}$ & $0.56^{0.86}_{0.26}$ & 1.0\% & - \\[5pt]

   4C 23.56$^{4}$ & 2.48 & 0.035 & 41.8 & 28 & 5.16 & $4.3^{+5.3}_{-2.6}$ & $0.8^{0.97}_{0.44}$ & $0.47^{0.72}_{0.26}$ & 55\% & 14.557 \\[5pt]

   MRC 0052-241$^{1,5}$ & 2.86 & 0.054 & 55.6 & 49 & 7.32 & $2^{+0.5}_{-0.4}$ & $0.94^{1.0}_{0.62}$ & $0.34^{0.59}_{0.18}$ & 55\% & 14.497 \\[5pt]

   MRC 0943-242$^{1,5}$ & 2.92 & 0.056 & 56.4 & 49 & 7.39 & $2.2^{+0.9}_{-0.7}$ & $0.94^{1.0}_{0.63}$ & $0.34^{0.58}_{0.18}$ & 55\% & 14.430 \\[5pt]

   SSA22-FLD$^{6,7,8}$ & 3.09 & 0.066 & 62.5 & 81 & 9.74 & $5^{+2}_{-2}$ & $1.0^{1.0}_{0.83}$ & $0.21^{0.44}_{0.11}$ & 29\% & - \\[5pt]

   MRC 0316-257$^{1,5}$ & 3.13 & 0.049 & 45.8 & 49 & 7.62 & $2.3^{+0.5}_{-0.4}$ & $0.95^{1.0}_{0.65}$ & $0.37^{0.62}_{0.20}$ & 59\% & 14.486 \\[5pt]

  TN J2009-3040$^{1,5}$ & 3.16 & 0.049 & 45.3 & 49 & 7.65 & $0.7^{+0.8}_{-0.6}$ & $0.95^{1.0}_{0.65}$ & $0.37^{0.62}_{0.20}$ & 2.4\% & - \\[5pt]

  TN J1338-1942$^{1,5,9}$ & 4.11 & 0.049 & 33.5 & 49 & 8.52 & $3.7^{+1.0}_{-0.8}$ & $0.97^{1.0}_{0.70}$ & $0.43^{0.70}_{0.23}$ & 71\% & 14.729 \\[5pt]

  TN J0924-2201$^{10}$ & 5.19 & 0.073 & 37.6 & 49 & 9.25 & $1.5^{+1.6}_{-1.0}$ & $0.98^{1.0}_{0.73}$ & $0.40^{0.68}_{0.21}$ & 30\% & - \\[5pt]

  SXDF-Object `A'$^{11}$ & 5.7 & 0.099 & 45.3 & 36 & 8.18 & $3.3^{+0.9}_{-0.9}$ & $0.94^{1.0}_{0.63}$ & $0.44^{0.72}_{0.23}$ & 79\% & 14.651 \\[5pt]

  SDF-12$^{3}$ & 6.01 & 0.05 & 21.4 & 36 & 8.31 & $16^{+7}_{-7}$ & $0.95^{1.0}_{0.64}$ & $0.62^{0.87}_{0.36}$ & 100\% & > 15.3 \\ [1ex]
   \hline
  \end{tabular}
\end{table*}

\begin{table*}
\caption{Estimated protocluster probabilities for the 12 strongest candidates from the CCPC catalogue \protect\citep{franck_candidate_2016}.
\newline \textbf{Notes:} (a) Redshift. (b) Measured galaxy overdensity within a cylindrical aperture with radius $R = 10 \mathrm{cMpc}$, and depth $2 \, \sigma_{z} = D$. (c) Full width redshift uncertainty. (c) Aperture length corresponding to redshift uncertainty. (d) Observation window area in square arc minutes. (d,e) Mean completeness and purity for each selection, and $5^{\mathrm{th}} - 95^{\mathrm{th}}$ percentile range. We use the lower percentile as our value for $C_{\mathrm{lim}}$ and $P_{\mathrm{lim}}$. (f) Protocluster probabilites from \protect\cite{franck_candidate_2016}, calculated using Figure 8 from \protect\cite{chiang_ancient_2013} using the same selection ($\mathrm{S_{S10}}$) (g) Derived protocluster probabilities, combining the \Proto\ and \PProto\ selections. (h) Descendant masses estimated using our fitting procedure. (i) Coefficient of determination.
\newline
\textbf{References}: (1) \protect\cite{venemans_protoclusters_2007} (2) \protect\cite{moller_detection_2001} (3) \protect\cite{steidel_large_1998} (4) \protect\cite{ellison_imaging_2001} }
\label{tab:obs}
\centering.
\begin{tabular}{|| l | c | c | c | c | c | c | c | c | c | c | c ||}
 \hline \hline
 Name & $z$ $^{\mathrm{a}}$ & $\delta_{g}$ $^{\mathrm{b}}$ &  $\sigma_{z}$ $^{\mathrm{c}}$ & $D \; (\mathrm{cMpc})$ $^{\mathrm{d}}$ & $C_{\mathrm{lim}}$ $^{\mathrm{e}}$ & $P_{\mathrm{lim}}$ $^{\mathrm{f}}$ & $P_{C} \; (\mathrm{F\&M})$ $^{\mathrm{g}}$ & $P_{C}({\mathrm{S_S10}})$ $^{\mathrm{h}}$ & $\mathrm{log_{10}}(M_{z=0} / M_{\odot})$ $^{\mathrm{i}}$ & $R^{2}$ $^{\mathrm{j}}$ \\
 \hline
 CCPC-z27-002 & 2.772 & $11.02 \pm 6.9$ & 0.007 & 14.9 & $1.0^{1.0}_{0.8}$ & $0.89^{1.0}_{0.54}$ & 100\% & 75\% & 14.47 & 0.63 \\[5pt]
CCPC-z29-001 & 2.918 & $ 11.21 \pm 4.76$ & 0.005 & 10.08 & $1.0^{1.0}_{0.67}$ & $1.0^{1.0}_{0.64}$ & 100\% & 46\% & 14.28 & 0.63 \\[5pt]
CCPC-z29-002$^{1}$ & 2.919 & $12.91 \pm 4.55$ & 0.009 & 18.12 & $1.0^{1.0}_{0.82}$ & $0.86^{1.0}_{0.5}$ & 100\% & 83\% & 14.67 & 0.61 \\[5pt]
CCPC-z30-001$^{2}$ & 3.035 & $18.78 \pm 10.14$ & 0.005 & 9.64 & $1.0^{1.0}_{0.67}$ & $1.0^{1.0}_{0.67}$ & 100\% & 74\% & 14.61 & 0.61 \\[5pt]
CCPC-z30-003$^{3}$ & 3.096 & $12.28 \pm 2.42$ & 0.008 & 15.10 & $1.0^{1.0}_{0.8}$ & $0.89^{1.0}_{0.55}$ & 100\% & 74\% & 14.55 & 0.63 \\[5pt]
CCPC-z31-003$^{1}$ & 3.133 & $9.80 \pm 2.77$ & 0.008 & 14.92 & $1.0^{1.0}_{0.8}$ & $0.89^{1.0}_{0.55}$ & 100\% & 48\% & 14.39 & 0.63 \\[5pt]
CCPC-z31-004 & 3.146 & $7.59 \pm 4.65$ & 0.006 & 11.14 & $1.0^{1.0}_{0.71}$ & $1.0^{1.0}_{0.62}$ & 85\% & 14\% & 14.09 & 0.63 \\[5pt]
CCPC-z31-005$^{1}$ & 3.152 & $17.77 \pm 9.19$ & 0.007 & 12.96 & $1.0^{1.0}_{0.75}$ & $0.92^{1.0}_{0.58}$ & 100\% & 86\% & 14.72 & 0.64 \\[5pt]
CCPC-z32-002 & 3.234 & $13.11 \pm 8.63$ & 0.003 & 5.40 & $0.8^{1.0}_{0.3}$ & $1.0^{1.0}_{0.67}$ & 100\% & 24\% & 14.11 & 0.49 \\[5pt]
CCPC-z33-002$^{4}$ & 3.372 & $7.44 \pm 4.47$ & 0.008 & 13.74 & $1.0^{1.0}_{0.78}$ & $0.91^{1.0}_{0.57}$ & 85\% & 42\% & 14.17 & 0.63 \\[5pt]
CCPC-z35-001 & 3.597 & $10.18 \pm 8.05$ & 0.003 & 4.80 & $0.6^{1.0}_{0.22}$ & $1.0^{1.0}_{0.67}$ & 100\% & 1\% & 13.80 & 0.32 \\[5pt]
CCPC-z36-001 & 3.644 & $23.50 \pm 14.39$ & 0.003 & 4.72 & $0.6^{1.0}_{0.2}$ & $1.0^{1.0}_{0.67}$ & 100\% & 72\% & 14.12 & 0.31 \\[1ex]
 \hline
\end{tabular}
\end{table*}

\begin{figure*}
	\includegraphics[width=\textwidth]{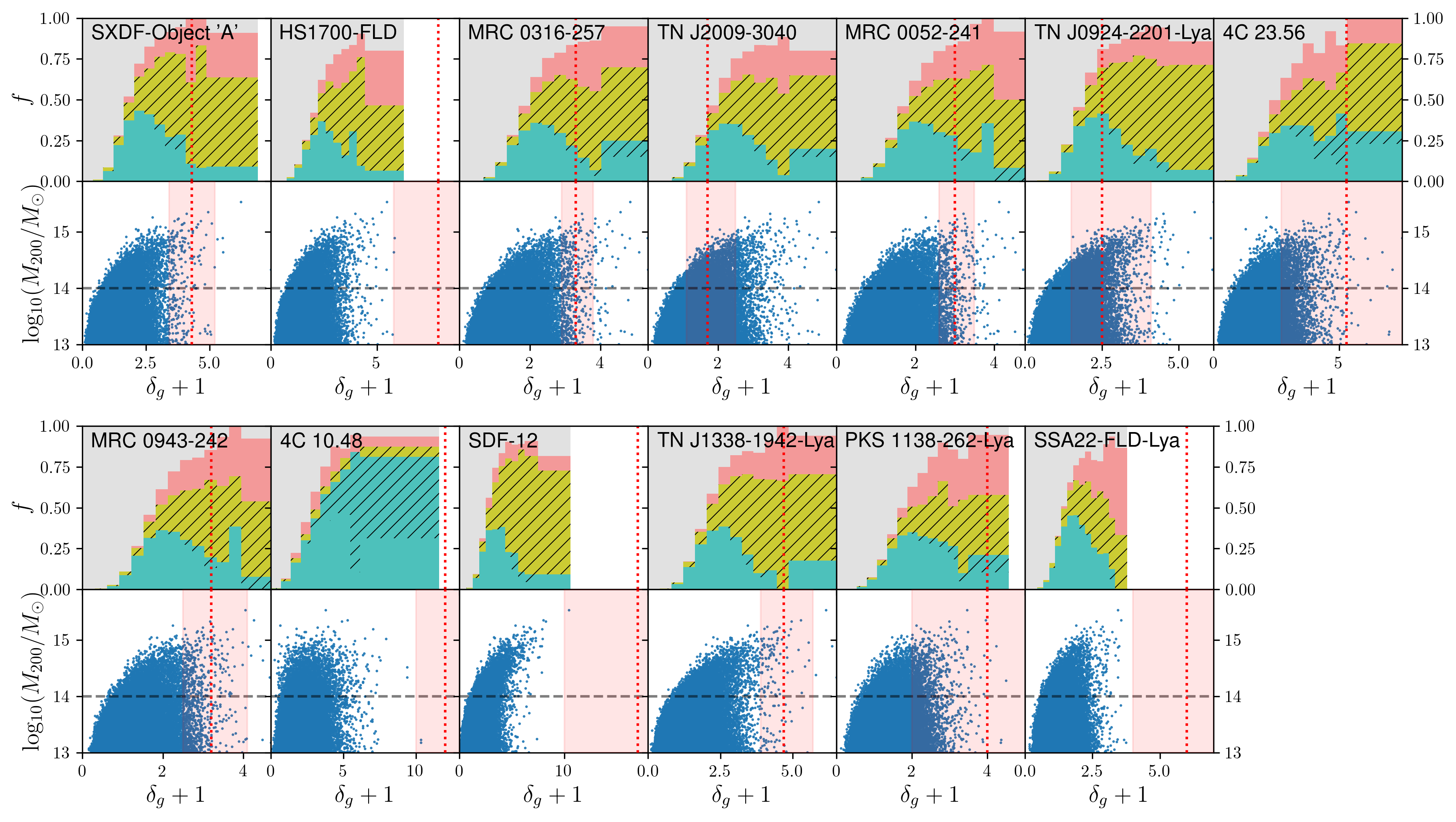}
    \caption{\textit{Top panels:} Probability distributions for each candidate from \Tab{obs_chiang} (labelled) for $100\,000$ random regions with the same dimensions as the given candidate. Probabilities are labelled identically to \Fig{probabilities_A}. The observationally measured overdensity is shown as a vertical dotted red line; where the overdensity exceeds the maximum overdensity from the random sampling, we show white space.
    \textit{Bottom panels:} Descendant mass against overdensity measured in the candidate aperture for all halos with $M / M_{\odot} > 10^{13}$. The cluster mass threshold is shown as the horizontal black dashed line. Uncertainties in the observationally measured overdensity are shaded in red.}
    \label{fig:obs_od_chiang}
\end{figure*}

\begin{figure*}
	\includegraphics[width=\textwidth]{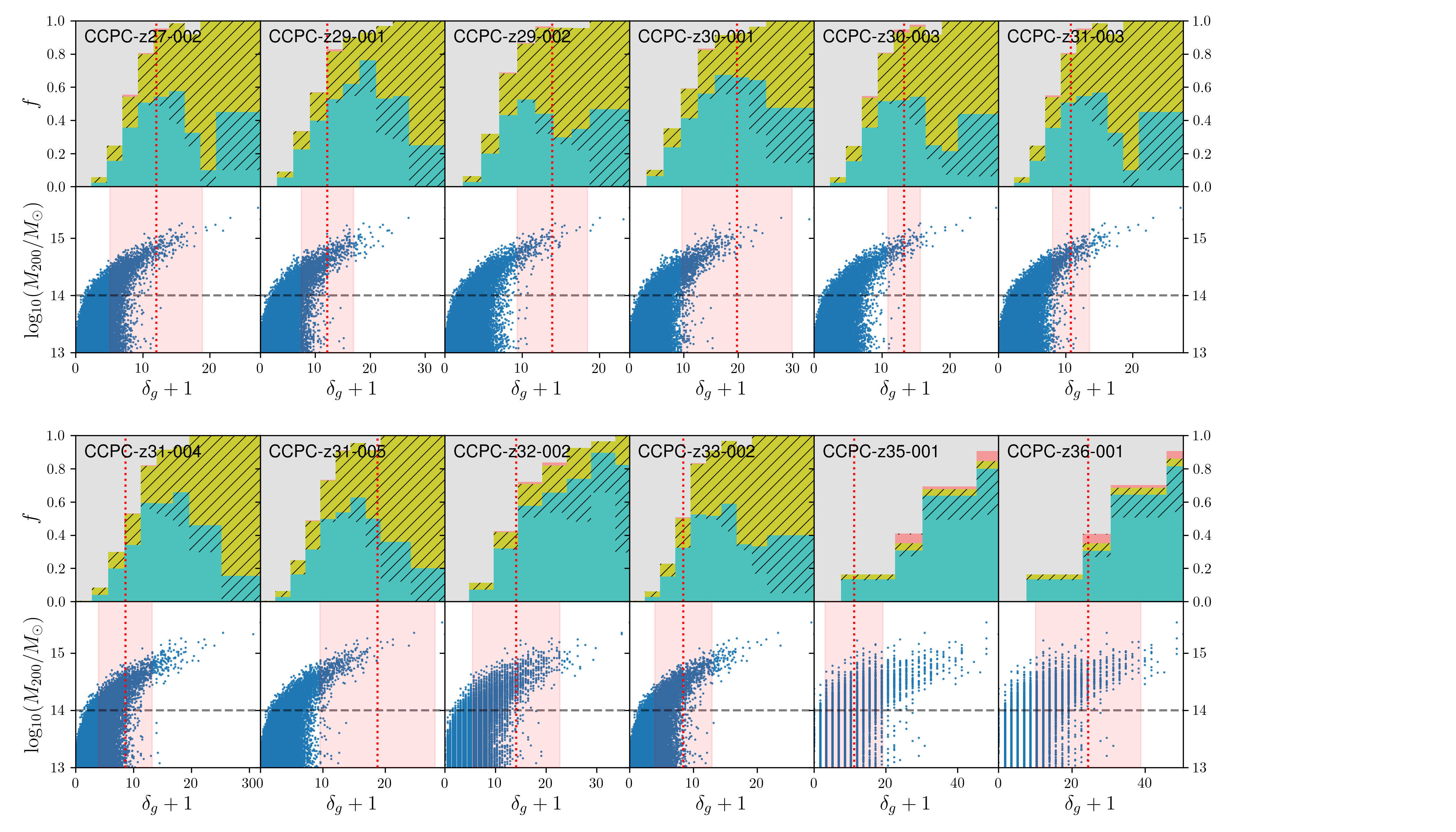}
    \caption{As for \Fig{obs_od_chiang}, but for the first 12 candidates from the Candidate Cluster and Protocluster Catalogue (CCPC) \citep{franck_candidate_2016} listed in \Tab{obs} and discussed in \Sec{discussion}.}
    \label{fig:obs_od}
\end{figure*}

\section{Discussion}
\label{sec:discussion}

In \Sec{overdensities} we presented an improved procedure for predicting the fate of observed galaxy overdensities.
To demonstrate, we apply the technique to a number of observational candidates in the literature.
\Tab{obs_chiang} lists estimated protocluster probabilities and descendant masses for 13 protocluster candidates from the literature, each of which have been studied in \cite{chiang_ancient_2013}.
We also apply the technique to the first 12 candidates presented in the Candidate Cluster and Protocluster Catalogue (CCPC) compiled in \cite{franck_candidate_2016}, shown in \Tab{obs}; this catalogue, whilst heterogenously selected, uses smaller, regular ($2R \sim D_{C}$) apertures to measure overdensity, and provides predictions for the protocluster probability and decendant mass derived from \cite{chiang_ancient_2013} that facilitate a direct comparison with our method.
In both cases we use an aperture with the same dimensions as the observations.\footnote{where rectangular apertures are used, we approximate with a cylinder of equal volume}
For the candidates in \Tab{obs_chiang} we use the \SelSFR\ selection, since all of these candidate overdensities are measured with star forming galaxies, whereas for \Tab{obs} we use the \SelTen\ selection identical to that used in \cite{franck_candidate_2016}; they acknowledge that this selection does not correspond exactly with the selection used to identify their candidates, but represents a conservative lower estimate (if the selection does include lower mass galaxies this would boost the overdensity measurement, and therefore the corresponding probabilities)
Each candidate is classified according to the $\mathrm{5^{th}}$ percentile of the completeness and purity of the protocluster population.

Many of the candidates in \Tab{obs_chiang} are measured with large apertures ($> \,(30 \, \mathrm{cMpc})^{3}$), which has a significant effect on derived descendant properties.
The bottom panels of \Fig{obs_od_chiang} show the relationship between overdensity and descendant mass for all halos with $M / M_{\odot} > 10^{13}$ in our model for the same aperture as each of these candidates; it is clear that for many it is very difficult to distinguish the protocluster population from the field in overdensity space.
$\mathrm{4C 10.48}$ is measured within a particularly pathological aperture ($R \ll D_{C}$) that leads to almost no distinction between the populations.
This effect can also be seen in the probability distributions in the top panels of \Fig{obs_od_chiang}.
Above intermediate overdensities the \Proto\ probability actually \textit{decreases} relative to the \PProto\ probability; if a large aperture happens to capture parts of two protoclusters, the overdensity will be boosted by both overdensities but the probabilities will be affected by the low completeness of each protocluster.

The measured overdensity for $\mathrm{4C 10.48}$ is much larger than that seen in randomly sampled regions or surrounding protoclusters, and we see similarly high overdensities for $\mathrm{HS1700-FLD}$, $\mathrm{SSA22-FLD-Ly\alpha}$ and $\mathrm{SDF-12}$.
We attribute these high overdensities to two primary effects.
First, each of these candidates is measured within a large aperture, which can be susceptible to aperture effects; our approach cannot distinguish the capture of two protoclusters within an aperture, or the chance alignment along a filamentary structure that is not destined to fall within the virial radius of the cluster at $z=0$.
Second, the selection criteria is not identical to that used for each candidate; a more conservative selection criteria could lead to a substantial boost in overdensity measurement \citep{chiang_ancient_2013}.
\cite{chiang_ancient_2013} note that $\mathrm{TN \; J2009-3040}$ is most likely a large group or low mass protocluster, and we come to a similar conclusion; \Fig{obs_od_chiang} shows that, whilst a number of protoclusters have a similar overdensity, a large number of groups also exhibit similar overdensities, which is reflected in the protocluster probability.

\Fig{obs_od} shows the probability and descendant mass distributions for the CCPC candidate apertures, listed in \Tab{obs} with probabilities and descendant mass estimates.
These candidates are typically measured with smaller apertures, which leads to greater distinction between protoclusters and the field, and high protocluster probabilities for sufficiently high overdensities; the majority are confirmed as protoclusters with high confidence.
CCPC-z32-002 is assigned a lower protocluster probability since it lies close to the overdensity threshold below which protoclusters are difficult to distinguish, and CCPC-z35-001 is ruled out with high confidence; whilst there are protoclusters with the same overdensity, the vast majority of objects with this overdensity have relatively low halo masses.

All of our results are simulation dependent, though we note that the pipeline is not, so it can be run again using catalogues from other simulations.
We also include protocluster regions in our calculation of the average field overdensity, so the field overdensity is an overestimate.
However, typical observable measures of field overdensity use the region in the foreground and background of the protocluster as a proxy for the `field' \citep{franck_candidate_2016, franck_candidate_2016-1}; since protoclusters have no sharp edge (see \Fig{compur_masscomp_dgal}), this approach may inadvertantly sample the protocluster overdensity tail, boosting the `field' overdensity. It's unclear to what degree these two effects cancel out.

\section{Summary}
\label{sec:summary}

We have used \textsc{L-Galaxies} to investigate the characteristics of galaxy protoclusters. Our findings are as follows:

\begin{itemize}
\setlength\itemsep{1em}

\item The completeness and purity of the protocluster galaxy population are maximised ($> 85\,\%$) at a radius of $\RC \approx 10\pm2 \, \mathrm{cMpc}$. This scale is insensitive to redshift and galaxy selections. Galaxy overdensities measured on $\RC$ provide high discrimination between protoclusters and the field, particularly at high redshift, and overdensities surrounding quasars and HzRGs are also best measured at $\RC$ since AGN are not centrally located within protoclusters.

\item Protocluster galaxies exhibit aspherical, prolate distributions, though this has little effect on their completeness and purity as measured within $\RC$ due to the lower density of galaxies in the field on their outskirts. Redshift space distortions slightly boost the measured overdensity, since protocluster galaxies tend to be infalling due to the Kaiser effect.

\item Using AGN as tracers at $z\gtrsim5$ is accurate but highly incomplete. The most luminous quasars at $z \sim 6$ are correlated with protocluster regions, but there are too few of them to act as tracers.

\item The most massive galaxies at all epochs preferentially appear in protocluster environments, and we see indirect evidence for the emergence of a red sequence in protoclusters through their greater asphericity and steeper completeness curves at $z \leqslant 3$.

\item We have demonstrated a procedure for generating protocluster probabilities based on their measured galaxy overdensity that can be applied to irregular apertures. We apply it to a range of redshifts and selection criteria, and provide fits between overdensity and descendant cluster mass. Low mass protoclusters cannot be discriminated due to overlap in overdensity space with field regions.

\end{itemize}

We make all of the code used in this paper public, at \url{https://github.com/christopherlovell/goa}.
It can be used to run the pipeline outlined in \Sec{gods}; we hope it will be of use to observers wishing to identify and characterise high-$z$ galaxy overdensities.

\section*{Acknowledgements}
The authors would like to thank the anonymous referee for useful comments, Daniel Cunnama, Romeel Dav\'e and Kate Storey-Fisher for encouraging discussions, and Yi-Kuan Chiang and Roderik Overzier for their helpful correspondence clarifying the overdensity measurement procedure in \cite{chiang_ancient_2013}. We also wish to acknowledge the use of the following open source software packages used in the analysis: Scipy \citep{scipy}, Astropy \citep{astropy} and Jupyter \citep{PER-GRA:2007}.
CCL acknowledges the support of a PhD studentship from the Science and Technology Facilities Council (STFC, grant number ST/P000252/1). 
PAT acknowledges support from the STFC (grant number ST/P000252/1). 
The Millennium Simulation was carried out by the Virgo Supercomputing Consortium at the Computing Centre of the Max-Planck Society in Garching.
The halo merger trees used are publicly available through the German Astronomical Virtual Observatory (GAVO) interface.\footnote{Available at http://www.mpa-garching.mpg.de/millennium/}




\bibliographystyle{mnras}
\bibliography{protoclusters,extra} 

\begin{thebibliography}{}
\makeatletter
\relax
\def\mn@urlcharsother{\let\do\@makeother \do\$\do\&\do\#\do\^\do\_\do\%\do\~}
\def\mn@doi{\begingroup\mn@urlcharsother \@ifnextchar [ {\mn@doi@}
  {\mn@doi@[]}}
\def\mn@doi@[#1]#2{\def\@tempa{#1}\ifx\@tempa\@empty \href
  {http://dx.doi.org/#2} {doi:#2}\else \href {http://dx.doi.org/#2} {#1}\fi
  \endgroup}
\def\mn@eprint#1#2{\mn@eprint@#1:#2::\@nil}
\def\mn@eprint@arXiv#1{\href {http://arxiv.org/abs/#1} {{\tt arXiv:#1}}}
\def\mn@eprint@dblp#1{\href {http://dblp.uni-trier.de/rec/bibtex/#1.xml}
  {dblp:#1}}
\def\mn@eprint@#1:#2:#3:#4\@nil{\def\@tempa {#1}\def\@tempb {#2}\def\@tempc
  {#3}\ifx \@tempc \@empty \let \@tempc \@tempb \let \@tempb \@tempa \fi \ifx
  \@tempb \@empty \def\@tempb {arXiv}\fi \@ifundefined
  {mn@eprint@\@tempb}{\@tempb:\@tempc}{\expandafter \expandafter \csname
  mn@eprint@\@tempb\endcsname \expandafter{\@tempc}}}

\bibitem[\protect\citeauthoryear{Adams et~al.,}{Adams
  et~al.}{2011}]{adams_hetdex_2011}
Adams J.~J.,  et~al., 2011, \mn@doi [ApJS] {10.1088/0067-0049/192/1/5}, 192, 5

\bibitem[\protect\citeauthoryear{Adams, Martini, Croxall, Overzier  \&
  Silverman}{Adams et~al.}{2015}]{adams_discovery_2015}
Adams S.~M.,  Martini P.,  Croxall K.~V.,  Overzier R.~A.,   Silverman J.~D.,
  2015, \mn@doi [MNRAS] {10.1093/mnras/stv065}, 448, 1335

\bibitem[\protect\citeauthoryear{Angulo \& White}{Angulo \&
  White}{2010}]{angulo_one_2010}
Angulo R.~E.,  White S. D.~M.,  2010, \mn@doi [MNRAS]
  {10.1111/j.1365-2966.2010.16459.x}, 405, 143

\bibitem[\protect\citeauthoryear{Angulo, Springel, White, Cole, Jenkins, Baugh
  \& Frenk}{Angulo et~al.}{2012}]{angulo_journey_2012}
Angulo R.~E.,  Springel V.,  White S. D.~M.,  Cole S.,  Jenkins A.,  Baugh
  C.~M.,   Frenk C.~S.,  2012, \mn@doi [MNRAS]
  {10.1111/j.1365-2966.2012.21783.x}, 425, 2722

\bibitem[\protect\citeauthoryear{{Astropy Collaboration} et~al.,}{{Astropy
  Collaboration} et~al.}{2013}]{astropy}
{Astropy Collaboration} et~al., 2013, \mn@doi [\aap]
  {10.1051/0004-6361/201322068}, \href
  {http://adsabs.harvard.edu/abs/2013A%26A...558A..33A} {558, A33}

\bibitem[\protect\citeauthoryear{Bett, Eke, Frenk, Jenkins, Helly  \&
  Navarro}{Bett et~al.}{2007}]{bett_spin_2007}
Bett P.,  Eke V.,  Frenk C.~S.,  Jenkins A.,  Helly J.,   Navarro J.,  2007,
  \mn@doi [MNRAS] {10.1111/j.1365-2966.2007.11432.x}, 376, 215

\bibitem[\protect\citeauthoryear{Bhattacharyya}{Bhattacharyya}{1946}]{bhattacharyya_measure_1946}
Bhattacharyya A.,  1946, Sankhya: The Indian Journal of Statistics (1933-1960),
  7, 401

\bibitem[\protect\citeauthoryear{{Cai} et~al.,}{{Cai}
  et~al.}{2016}]{Cai_mapping_2016}
{Cai} Z.,  et~al., 2016, \mn@doi [\apj] {10.3847/1538-4357/833/2/135}, \href
  {http://adsabs.harvard.edu/abs/2016ApJ...833..135C} {833, 135}

\bibitem[\protect\citeauthoryear{Capak et~al.,}{Capak
  et~al.}{2011}]{capak_massive_2011}
Capak P.~L.,  et~al., 2011, \mn@doi [Nature] {10.1038/nature09681}, 470, 233

\bibitem[\protect\citeauthoryear{Casey, Narayanan  \& Cooray}{Casey
  et~al.}{2014}]{casey_dusty_2014}
Casey C.~M.,  Narayanan D.,   Cooray A.,  2014, \mn@doi [Phys. Rep.]
  {10.1016/j.physrep.2014.02.009}, 541, 45

\bibitem[\protect\citeauthoryear{Chiang, Overzier  \& Gebhardt}{Chiang
  et~al.}{2013}]{chiang_ancient_2013}
Chiang Y.-K.,  Overzier R.,   Gebhardt K.,  2013, \mn@doi [ApJ]
  {10.1088/0004-637X/779/2/127}, 779, 127

\bibitem[\protect\citeauthoryear{Chiang, Overzier  \& Gebhardt}{Chiang
  et~al.}{2014}]{chiang_discovery_2014}
Chiang Y.-K.,  Overzier R.,   Gebhardt K.,  2014, \mn@doi [ApJL]
  {10.1088/2041-8205/782/1/L3}, 782, L3

\bibitem[\protect\citeauthoryear{Chiang, Overzier, Gebhardt  \&
  Henriques}{Chiang et~al.}{2017}]{chiang_galaxy_2017}
Chiang Y.-K.,  Overzier R.~A.,  Gebhardt K.,   Henriques B.,  2017,
  arXiv:1705.01634 [astro-ph]

\bibitem[\protect\citeauthoryear{Clay, Thomas, Wilkins  \& Henriques}{Clay
  et~al.}{2015}]{clay_galaxy_2015}
Clay S.,  Thomas P.,  Wilkins S.,   Henriques B.,  2015, \mn@doi [MNRAS]
  {10.1093/mnras/stv818}, 451, 2692

\bibitem[\protect\citeauthoryear{Contini, Lucia, Hatch, Borgani  \&
  Kang}{Contini et~al.}{2016}]{contini_semi-analytic_2016}
Contini E.,  Lucia G.~D.,  Hatch N.,  Borgani S.,   Kang X.,  2016, \mn@doi
  [MNRAS] {10.1093/mnras/stv2852}, 456, 1924

\bibitem[\protect\citeauthoryear{Cooke, Hatch, Muldrew, Rigby  \& Kurk}{Cooke
  et~al.}{2014}]{cooke_z_2014}
Cooke E.~A.,  Hatch N.~A.,  Muldrew S.~I.,  Rigby E.~E.,   Kurk J.~D.,  2014,
  \mn@doi [MNRAS] {10.1093/mnras/stu522}, 440, 3262

\bibitem[\protect\citeauthoryear{Croton et~al.,}{Croton
  et~al.}{2006}]{croton_many_2006}
Croton D.~J.,  et~al., 2006, \mn@doi [MNRAS]
  {10.1111/j.1365-2966.2005.09675.x}, 365, 11

\bibitem[\protect\citeauthoryear{Diener et~al.,}{Diener
  et~al.}{2015}]{diener_protocluster_2015}
Diener C.,  et~al., 2015, \mn@doi [ApJ] {10.1088/0004-637X/802/1/31}, 802, 31

\bibitem[\protect\citeauthoryear{Dressler}{Dressler}{1980}]{dressler_galaxy_1980}
Dressler A.,  1980, \mn@doi [ApJ] {10.1086/157753}, 236, 351

\bibitem[\protect\citeauthoryear{Dunlop \& Peacock}{Dunlop \&
  Peacock}{1990}]{dunlop_redshift_1990}
Dunlop J.~S.,  Peacock J.~A.,  1990, Monthly Notices of the Royal Astronomical
  Society, 247, 19

\bibitem[\protect\citeauthoryear{Ellison, Pettini, Steidel  \& Shapley}{Ellison
  et~al.}{2001}]{ellison_imaging_2001}
Ellison S.~L.,  Pettini M.,  Steidel C.~C.,   Shapley A.~E.,  2001, \mn@doi
  [ApJ] {10.1086/319457}, 549, 770

\bibitem[\protect\citeauthoryear{Fanidakis, Baugh, Benson, Bower, Cole, Done
  \& Frenk}{Fanidakis et~al.}{2011}]{fanidakis_grand_2011}
Fanidakis N.,  Baugh C.~M.,  Benson A.~J.,  Bower R.~G.,  Cole S.,  Done C.,
  Frenk C.~S.,  2011, \mn@doi [MNRAS] {10.1111/j.1365-2966.2010.17427.x}, 410,
  53

\bibitem[\protect\citeauthoryear{F{\'e}vre, Deltorn, Crampton  \&
  Dickinson}{F{\'e}vre et~al.}{1996}]{fevre_clustering_1996}
F{\'e}vre O.~L.,  Deltorn J.~M.,  Crampton D.,   Dickinson M.,  1996, \mn@doi
  [ApJ] {10.1086/310319}, 471, L11

\bibitem[\protect\citeauthoryear{F{\'e}vre et~al.,}{F{\'e}vre
  et~al.}{2015}]{fevre_vimos_2015}
F{\'e}vre O.~L.,  et~al., 2015, \mn@doi [A\&A] {10.1051/0004-6361/201423829},
  576, A79

\bibitem[\protect\citeauthoryear{Franck \& McGaugh}{Franck \&
  McGaugh}{2016a}]{franck_candidate_2016}
Franck J.~R.,  McGaugh S.~S.,  2016a, arXiv:1610.00713 [astro-ph]

\bibitem[\protect\citeauthoryear{Franck \& McGaugh}{Franck \&
  McGaugh}{2016b}]{franck_candidate_2016-1}
Franck J.~R.,  McGaugh S.~S.,  2016b, \mn@doi [ApJ]
  {10.3847/0004-637X/817/2/158}, 817, 158

\bibitem[\protect\citeauthoryear{Franx, Illingworth  \& de Zeeuw}{Franx
  et~al.}{1991}]{franx_ordered_1991}
Franx M.,  Illingworth G.,   de Zeeuw T.,  1991, \mn@doi [ApJ]
  {10.1086/170769}, 383, 112

\bibitem[\protect\citeauthoryear{Galametz et~al.,}{Galametz
  et~al.}{2010}]{galametz_galaxy_2010}
Galametz A.,  et~al., 2010, \mn@doi [A\&A] {10.1051/0004-6361/201015035}, 522,
  A58

\bibitem[\protect\citeauthoryear{Guo et~al.,}{Guo
  et~al.}{2011}]{guo_dwarf_2011}
Guo Q.,  et~al., 2011, \mn@doi [MNRAS] {10.1111/j.1365-2966.2010.18114.x}, 413,
  101

\bibitem[\protect\citeauthoryear{Hatch et~al.,}{Hatch
  et~al.}{2011a}]{hatch_galaxy_2011}
Hatch N.~A.,  et~al., 2011a, \mn@doi [MNRAS]
  {10.1111/j.1365-2966.2010.17538.x}, 410, 1537

\bibitem[\protect\citeauthoryear{Hatch, Kurk, Pentericci, Venemans, Kuiper,
  Miley  \& Röttgering}{Hatch et~al.}{2011b}]{hatch_halpha_2011}
Hatch N.~A.,  Kurk J.~D.,  Pentericci L.,  Venemans B.~P.,  Kuiper E.,  Miley
  G.~K.,   Röttgering H. J.~A.,  2011b, \mn@doi [MNRAS]
  {10.1111/j.1365-2966.2011.18735.x}, 415, 2993

\bibitem[\protect\citeauthoryear{Hatch et~al.,}{Hatch
  et~al.}{2014}]{hatch_why_2014}
Hatch N.~A.,  et~al., 2014, \mn@doi [MNRAS] {10.1093/mnras/stu1725}, 445, 280

\bibitem[\protect\citeauthoryear{Hennawi, Prochaska, Cantalupo  \&
  Arrigoni-Battaia}{Hennawi et~al.}{2015}]{hennawi_quasar_2015}
Hennawi J.~F.,  Prochaska J.~X.,  Cantalupo S.,   Arrigoni-Battaia F.,  2015,
  \mn@doi [Science] {10.1126/science.aaa5397}, 348, 779

\bibitem[\protect\citeauthoryear{Henriques, White, Thomas, Angulo, Guo, Lemson,
  Springel  \& Overzier}{Henriques et~al.}{2015}]{henriques_galaxy_2015}
Henriques B. M.~B.,  White S. D.~M.,  Thomas P.~A.,  Angulo R.,  Guo Q.,
  Lemson G.,  Springel V.,   Overzier R.,  2015, \mn@doi [MNRAS]
  {10.1093/mnras/stv705}, 451, 2663

\bibitem[\protect\citeauthoryear{Hopkins, Richards  \& Hernquist}{Hopkins
  et~al.}{2007}]{hopkins_observational_2007}
Hopkins P.~F.,  Richards G.~T.,   Hernquist L.,  2007, \mn@doi [ApJ]
  {10.1086/509629}, 654, 731

\bibitem[\protect\citeauthoryear{Husband, Bremer, Stanway, Davies, Lehnert  \&
  Douglas}{Husband et~al.}{2013}]{husband_are_2013}
Husband K.,  Bremer M.~N.,  Stanway E.~R.,  Davies L. J.~M.,  Lehnert M.~D.,
  Douglas L.~S.,  2013, \mn@doi [MNRAS] {10.1093/mnras/stt642}, 432, 2869

\bibitem[\protect\citeauthoryear{Jarvis, Rawlings, Willott, Blundell, Eales  \&
  Lacy}{Jarvis et~al.}{2001}]{jarvis_redshift_2001}
Jarvis M.~J.,  Rawlings S.,  Willott C.~J.,  Blundell K.~M.,  Eales S.,   Lacy
  M.,  2001, \mn@doi [Monthly Notices of the Royal Astronomical Society]
  {10.1046/j.1365-8711.2001.04778.x}, 327, 907

\bibitem[\protect\citeauthoryear{Jones, Oliphant, Peterson  et~al.}{Jones
  et~al.}{2001}]{scipy}
Jones E.,  Oliphant T.,  Peterson P.,   et~al., 2001, {SciPy}: Open source
  scientific tools for {Python}, \url {http://www.scipy.org/}

\bibitem[\protect\citeauthoryear{Kaiser}{Kaiser}{1987}]{kaiser_clustering_1987}
Kaiser N.,  1987, \mn@doi [Monthly Notices of the Royal Astronomical Society]
  {10.1093/mnras/227.1.1}, 227, 1

\bibitem[\protect\citeauthoryear{Koyama, Kodama, Tadaki, Hayashi, Tanaka,
  Smail, Tanaka  \& Kurk}{Koyama et~al.}{2012}]{koyama_massive_2012}
Koyama Y.,  Kodama T.,  Tadaki K.-i.,  Hayashi M.,  Tanaka M.,  Smail I.,
  Tanaka I.,   Kurk J.,  2012, \mn@doi [MNRAS] {10.1093/mnras/sts133}

\bibitem[\protect\citeauthoryear{Lemaux et~al.,}{Lemaux
  et~al.}{2017}]{lemaux_vimos_2017}
Lemaux B.~C.,  et~al., 2017, arXiv:1703.10170 [astro-ph]

\bibitem[\protect\citeauthoryear{Lucia \& Blaizot}{Lucia \&
  Blaizot}{2007}]{lucia_hierarchical_2007}
Lucia G.~D.,  Blaizot J.,  2007, \mn@doi [MNRAS]
  {10.1111/j.1365-2966.2006.11287.x}, 375, 2

\bibitem[\protect\citeauthoryear{Matsuda et~al.,}{Matsuda
  et~al.}{2005}]{matsuda_large_2005}
Matsuda Y.,  et~al., 2005, \mn@doi [ApJ] {10.1086/499071}, 634, L125

\bibitem[\protect\citeauthoryear{Mazzucchelli, Ba{\~n}ados, Decarli, Farina,
  Venemans, Walter  \& Overzier}{Mazzucchelli
  et~al.}{2017}]{mazzucchelli_no_2017}
Mazzucchelli C.,  Ba{\~n}ados E.,  Decarli R.,  Farina E.~P.,  Venemans B.~P.,
  Walter F.,   Overzier R.,  2017, \mn@doi [The Astrophysical Journal]
  {10.3847/1538-4357/834/1/83}, 834, 83

\bibitem[\protect\citeauthoryear{Miley et~al.,}{Miley
  et~al.}{2006}]{miley_spiderweb_2006}
Miley G.~K.,  et~al., 2006, \mn@doi [ApJ] {10.1086/508534}, 650, L29

\bibitem[\protect\citeauthoryear{Miller, Chapman, Hayward, Behroozi, Bradford,
  Willott  \& Wagg}{Miller et~al.}{2016}]{miller_investigating_2016}
Miller T.~B.,  Chapman S.~C.,  Hayward C.~C.,  Behroozi P.~S.,  Bradford C.~M.,
   Willott C.~J.,   Wagg J.,  2016, arXiv:1611.08552 [astro-ph]

\bibitem[\protect\citeauthoryear{M{\o}ller \& Fynbo}{M{\o}ller \&
  Fynbo}{2001}]{moller_detection_2001}
M{\o}ller P.,  Fynbo J.~U.,  2001, \mn@doi [A\&A] {10.1051/0004-6361:20010606},
  372

\bibitem[\protect\citeauthoryear{Monaco, M{\o}ller, Fynbo, Weidinger, Ledoux
  \& Theuns}{Monaco et~al.}{2005}]{monaco_tracing_2005}
Monaco P.,  M{\o}ller P.,  Fynbo J. P.~U.,  Weidinger M.,  Ledoux C.,   Theuns
  T.,  2005, \mn@doi [A\&A] {10.1051/0004-6361:20042570}, 440

\bibitem[\protect\citeauthoryear{Morselli et~al.,}{Morselli
  et~al.}{2014}]{morselli_primordial_2014}
Morselli L.,  et~al., 2014, \mn@doi [A\&A] {10.1051/0004-6361/201423853}, 568,
  A1

\bibitem[\protect\citeauthoryear{Muldrew et~al.,}{Muldrew
  et~al.}{2012}]{muldrew_measures_2012}
Muldrew S.~I.,  et~al., 2012, \mn@doi [MNRAS]
  {10.1111/j.1365-2966.2011.19922.x}, 419, 2670

\bibitem[\protect\citeauthoryear{Muldrew, Hatch  \& Cooke}{Muldrew
  et~al.}{2015}]{muldrew_what_2015}
Muldrew S.~I.,  Hatch N.~A.,   Cooke E.~A.,  2015, \mn@doi [MNRAS]
  {10.1093/mnras/stv1449}, 452, 2528

\bibitem[\protect\citeauthoryear{Orsi, Fanidakis, Lacey  \& Baugh}{Orsi
  et~al.}{2016}]{orsi_environments_2016}
Orsi Ã.~A.,  Fanidakis N.,  Lacey C.~G.,   Baugh C.~M.,  2016, \mn@doi [MNRAS]
  {10.1093/mnras/stv2919}, 456, 3827

\bibitem[\protect\citeauthoryear{Ouchi et~al.,}{Ouchi
  et~al.}{2005}]{ouchi_discovery_2005}
Ouchi M.,  et~al., 2005, \mn@doi [ApJ] {10.1086/428499}, 620, L1

\bibitem[\protect\citeauthoryear{Overzier}{Overzier}{2016}]{overzier_realm_2016}
Overzier R.~A.,  2016, arXiv:1610.05201 [astro-ph]

\bibitem[\protect\citeauthoryear{Overzier, Guo, Kauffmann, Lucia, Bouwens  \&
  Lemson}{Overzier et~al.}{2009}]{overzier_cdm_2009}
Overzier R.~A.,  Guo Q.,  Kauffmann G.,  Lucia G.~D.,  Bouwens R.,   Lemson G.,
   2009, \mn@doi [MNRAS] {10.1111/j.1365-2966.2008.14264.x}, 394, 577

\bibitem[\protect\citeauthoryear{P\'erez \& Granger}{P\'erez \&
  Granger}{2007}]{PER-GRA:2007}
P\'erez F.,  Granger B.~E.,  2007, \mn@doi [Comput. Sci. Eng.]
  {10.1109/MCSE.2007.53}, 9, 21

\bibitem[\protect\citeauthoryear{{Planck Collaboration} et~al.,}{{Planck
  Collaboration} et~al.}{2014}]{planck_collaboration_planck_2014}
{Planck Collaboration} et~al., 2014, \mn@doi [A\&A]
  {10.1051/0004-6361/201321529}, 571, A1

\bibitem[\protect\citeauthoryear{Ramos~Almeida, Bessiere, Tadhunter, Inskip,
  Morganti, Dicken, Gonz{\'a}lez-Serrano  \& Holt}{Ramos~Almeida
  et~al.}{2013}]{ramos_almeida_environments_2013}
Ramos~Almeida C.,  Bessiere P.~S.,  Tadhunter C.~N.,  Inskip K.~J.,  Morganti
  R.,  Dicken D.,  Gonz{\'a}lez-Serrano J.~I.,   Holt J.,  2013, \mn@doi
  [MNRAS] {10.1093/mnras/stt1595}, 436, 997

\bibitem[\protect\citeauthoryear{Rigby, Best, Brookes, Peacock, Dunlop,
  R{\"o}ttgering, Wall  \& Ker}{Rigby
  et~al.}{2011}]{rigby_luminosity-dependent_2011}
Rigby E.~E.,  Best P.~N.,  Brookes M.~H.,  Peacock J.~A.,  Dunlop J.~S.,
  R{\"o}ttgering H. J.~A.,  Wall J.~V.,   Ker L.,  2011, \mn@doi [Monthly
  Notices of the Royal Astronomical Society]
  {10.1111/j.1365-2966.2011.19167.x}, 416, 1900

\bibitem[\protect\citeauthoryear{Schneider, Frenk  \& Cole}{Schneider
  et~al.}{2012}]{schneider_shapes_2012}
Schneider M.~D.,  Frenk C.~S.,   Cole S.,  2012, \mn@doi [JCAP]
  {10.1088/1475-7516/2012/05/030}, 2012, 030

\bibitem[\protect\citeauthoryear{Shattow, Croton, Skibba, Muldrew, Pearce  \&
  Abbas}{Shattow et~al.}{2013}]{shattow_measures_2013}
Shattow G.~M.,  Croton D.~J.,  Skibba R.~A.,  Muldrew S.~I.,  Pearce F.~R.,
  Abbas U.,  2013, \mn@doi [MNRAS] {10.1093/mnras/stt998}, 433, 3314

\bibitem[\protect\citeauthoryear{Shimakawa, Kodama, Tadaki, Tanaka, Hayashi  \&
  Koyama}{Shimakawa et~al.}{2014}]{shimakawa_identification_2014}
Shimakawa R.,  Kodama T.,  Tadaki K.-i.,  Tanaka I.,  Hayashi M.,   Koyama Y.,
  2014, \mn@doi [Monthly Notices of the Royal Astronomical Society: Letters]
  {10.1093/mnrasl/slu029}, 441, L1

\bibitem[\protect\citeauthoryear{{Shimakawa} et~al.,}{{Shimakawa}
  et~al.}{2018}]{shimakawa_2018}
{Shimakawa} R.,  et~al., 2018, \mn@doi [\mnras] {10.1093/mnras/stx2494}, \href
  {http://adsabs.harvard.edu/abs/2018MNRAS.473.1977S} {473, 1977}

\bibitem[\protect\citeauthoryear{Shimasaku et~al.,}{Shimasaku
  et~al.}{2003}]{shimasaku_subaru_2003}
Shimasaku K.,  et~al., 2003, \mn@doi [ApJ] {10.1086/374880}, 586, L111

\bibitem[\protect\citeauthoryear{Spergel et~al.,}{Spergel
  et~al.}{2003}]{spergel_first-year_2003}
Spergel D.~N.,  et~al., 2003, \mn@doi [ApJS] {10.1086/377226}, 148, 175

\bibitem[\protect\citeauthoryear{Spitler et~al.,}{Spitler
  et~al.}{2012}]{spitler_first_2012}
Spitler L.~R.,  et~al., 2012, \mn@doi [ApJL] {10.1088/2041-8205/748/2/L21},
  748, L21

\bibitem[\protect\citeauthoryear{Springel et~al.,}{Springel
  et~al.}{2005}]{springel_simulations_2005}
Springel V.,  et~al., 2005, \mn@doi [Nature] {10.1038/nature03597}, 435, 629

\bibitem[\protect\citeauthoryear{Steidel, Adelberger, Dickinson, Giavalisco,
  Pettini  \& Kellogg}{Steidel et~al.}{1998}]{steidel_large_1998}
Steidel C.~C.,  Adelberger K.~L.,  Dickinson M.,  Giavalisco M.,  Pettini M.,
  Kellogg M.,  1998, \mn@doi [ApJ] {10.1086/305073}, 492, 428

\bibitem[\protect\citeauthoryear{Steidel, Adelberger, Shapley, Pettini,
  Dickinson  \& Giavalisco}{Steidel et~al.}{2000}]{steidel_ly_2000}
Steidel C.~C.,  Adelberger K.~L.,  Shapley A.~E.,  Pettini M.,  Dickinson M.,
  Giavalisco M.,  2000, \mn@doi [ApJ] {10.1086/308568}, 532, 170

\bibitem[\protect\citeauthoryear{{Steidel}, {Adelberger}, {Shapley}, {Erb},
  {Reddy}  \& {Pettini}}{{Steidel} et~al.}{2005}]{steidel_spectroscopic_2005}
{Steidel} C.~C.,  {Adelberger} K.~L.,  {Shapley} A.~E.,  {Erb} D.~K.,  {Reddy}
  N.~A.,   {Pettini} M.,  2005, \mn@doi [\apj] {10.1086/429989}, \href
  {http://adsabs.harvard.edu/abs/2005ApJ...626...44S} {626, 44}

\bibitem[\protect\citeauthoryear{Strazzullo et~al.,}{Strazzullo
  et~al.}{2013}]{strazzullo_galaxy_2013}
Strazzullo V.,  et~al., 2013, \mn@doi [The Astrophysical Journal]
  {10.1088/0004-637X/772/2/118}, 772, 118

\bibitem[\protect\citeauthoryear{Suwa, Habe  \& Yoshikawa}{Suwa
  et~al.}{2006}]{suwa_protoclusters_2006}
Suwa T.,  Habe A.,   Yoshikawa K.,  2006, \mn@doi [ApJ] {10.1086/506607}, 646,
  L5

\bibitem[\protect\citeauthoryear{Tanaka et~al.,}{Tanaka
  et~al.}{2011}]{tanaka_discovery_2011}
Tanaka I.,  et~al., 2011, \mn@doi [Publications of the Astronomical Society of
  Japan] {10.1093/pasj/63.sp2.S415}, 63, 415

\bibitem[\protect\citeauthoryear{Thomas et~al.,}{Thomas
  et~al.}{1998}]{thomas_structure_1998}
Thomas P.~A.,  et~al., 1998, \mn@doi [MNRAS]
  {10.1046/j.1365-8711.1998.01491.x}, 296, 1061

\bibitem[\protect\citeauthoryear{Toshikawa et~al.,}{Toshikawa
  et~al.}{2012}]{toshikawa_discovery_2012}
Toshikawa J.,  et~al., 2012, \mn@doi [ApJ] {10.1088/0004-637X/750/2/137}, 750,
  137

\bibitem[\protect\citeauthoryear{Toshikawa et~al.,}{Toshikawa
  et~al.}{2016}]{toshikawa_systematic_2016}
Toshikawa J.,  et~al., 2016, \mn@doi [ApJ] {10.3847/0004-637X/826/2/114}, 826,
  114

\bibitem[\protect\citeauthoryear{Uchiyama et~al.,}{Uchiyama
  et~al.}{2017}]{uchiyama_luminous_2017}
Uchiyama H.,  et~al., 2017, arXiv:1704.06050 [astro-ph]

\bibitem[\protect\citeauthoryear{Venemans et~al.,}{Venemans
  et~al.}{2002}]{venemans_most_2002}
Venemans B.~P.,  et~al., 2002, \mn@doi [ApJ] {10.1086/340563}, 569, L11

\bibitem[\protect\citeauthoryear{Venemans et~al.,}{Venemans
  et~al.}{2004}]{venemans_discovery_2004}
Venemans B.~P.,  et~al., 2004, \mn@doi [Astronomy and Astrophysics]
  {10.1051/0004-6361:200400041}, 424, L17

\bibitem[\protect\citeauthoryear{Venemans et~al.,}{Venemans
  et~al.}{2005}]{venemans_properties_2005}
Venemans B.~P.,  et~al., 2005, \mn@doi [Astronomy and Astrophysics]
  {10.1051/0004-6361:20042038}, 431, 793

\bibitem[\protect\citeauthoryear{Venemans et~al.,}{Venemans
  et~al.}{2007}]{venemans_protoclusters_2007}
Venemans B.~P.,  et~al., 2007, \mn@doi [A\&A] {10.1051/0004-6361:20053941},
  461, 823

\bibitem[\protect\citeauthoryear{Vikhlinin, Kravtsov, Markevich, Sunyaev  \&
  Churazov}{Vikhlinin et~al.}{2014}]{vikhlinin_clusters_2014}
Vikhlinin A.~A.,  Kravtsov A.~V.,  Markevich M.~L.,  Sunyaev R.~A.,   Churazov
  E.~M.,  2014, \mn@doi [Phys.-Usp.] {10.3367/UFNe.0184.201404a.0339}, 57, 317

\bibitem[\protect\citeauthoryear{Vulcani et~al.,}{Vulcani
  et~al.}{2011}]{vulcani_galaxy_2011}
Vulcani B.,  et~al., 2011, \mn@doi [MNRAS] {10.1111/j.1365-2966.2010.17904.x},
  412, 246

\bibitem[\protect\citeauthoryear{Willott, Percival, McLure, Crampton,
  Hutchings, Jarvis, Sawicki  \& Simard}{Willott
  et~al.}{2005}]{willott_imaging_2005}
Willott C.~J.,  Percival W.~J.,  McLure R.~J.,  Crampton D.,  Hutchings J.~B.,
  Jarvis M.~J.,  Sawicki M.,   Simard L.,  2005, \mn@doi [ApJ]
  {10.1086/430168}, 626, 657

\bibitem[\protect\citeauthoryear{Wu, Hahn, Wechsler, Mao  \& Behroozi}{Wu
  et~al.}{2013}]{wu_rhapsody._2013}
Wu H.-Y.,  Hahn O.,  Wechsler R.~H.,  Mao Y.-Y.,   Behroozi P.~S.,  2013,
  \mn@doi [ApJ] {10.1088/0004-637X/763/2/70}, 763, 70

\bibitem[\protect\citeauthoryear{Wylezalek et~al.,}{Wylezalek
  et~al.}{2013}]{wylezalek_galaxy_2013}
Wylezalek D.,  et~al., 2013, \mn@doi [ApJ] {10.1088/0004-637X/769/1/79}, 769,
  79

\bibitem[\protect\citeauthoryear{Yamada, Nakamura, Matsuda, Hayashino,
  Yamauchi, Morimoto, Kousai  \& Umemura}{Yamada
  et~al.}{2012}]{yamada_panoramic_2012}
Yamada T.,  Nakamura Y.,  Matsuda Y.,  Hayashino T.,  Yamauchi R.,  Morimoto
  N.,  Kousai K.,   Umemura M.,  2012, \mn@doi [AJ]
  {10.1088/0004-6256/143/4/79}, 143, 79

\makeatother
\end{thebibliography}



\appendix

\section{Overdensity Statistics}
\label{sec:appendix1}

\begin{figure}
	\includegraphics[width=\columnwidth]{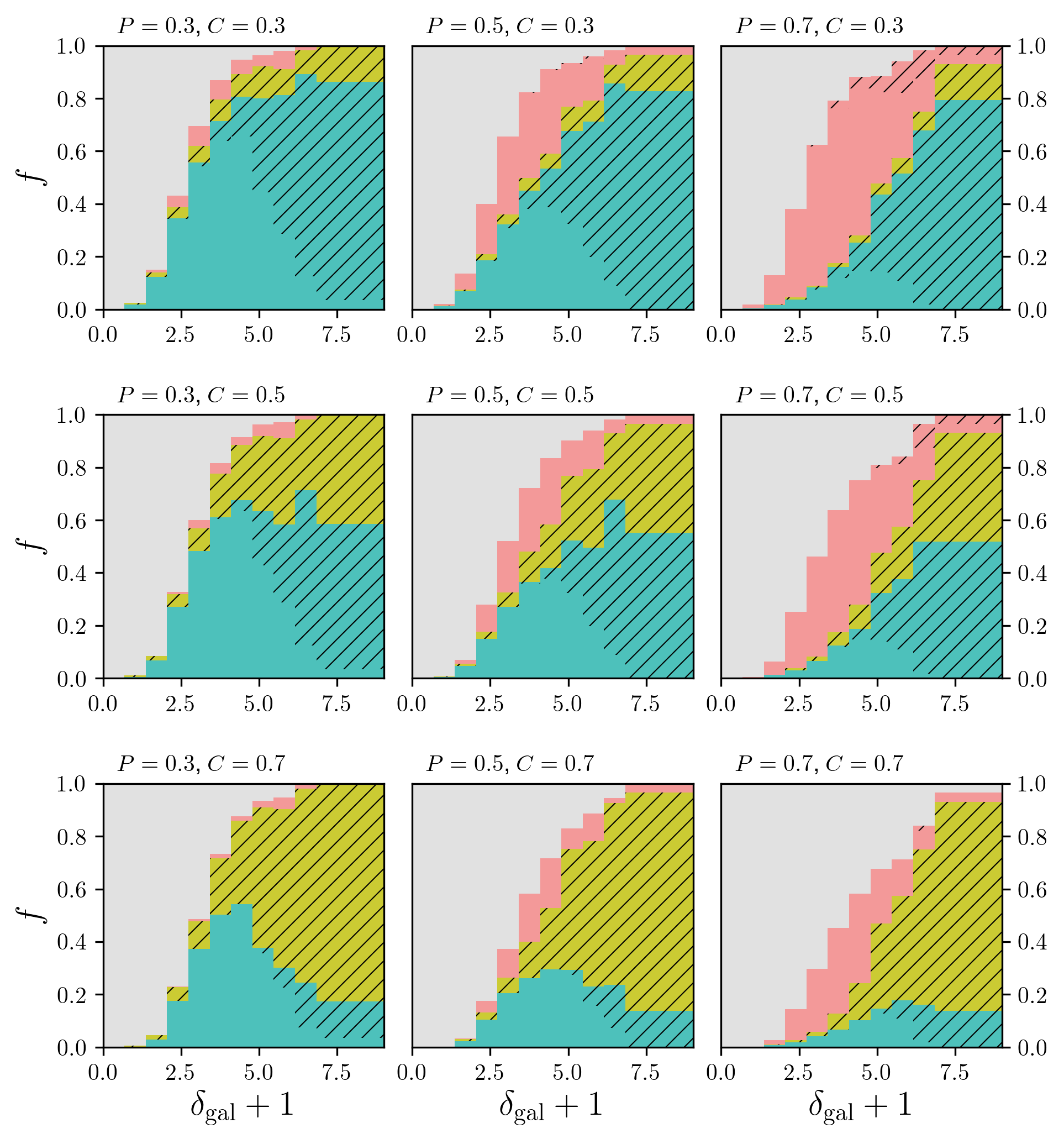}
    \caption{Fractional probability distributions for different choices of \Clim\ and \Plim\ (See \Fig{probabilities_A} for legend). In general, the higher the purity constraint, the more regions are classified as \ProtoF, and the higher the completeness constraint, the more regions are classified as \PProto. Higher \Plim\ can also lead to higher \Field\ probabilities.}
    \label{fig:probabilities_C}
\end{figure}

\Fig{probabilities_C} shows the effect of adjusting our free parameters, \Clim\ and \Plim, whilst keeping a fixed aperture volume ($R = D/2 = 10 \, \mathrm{cMpc}$).
Changing \Clim\ principally affects the ratio of probability of \PProto\ to \Proto, and \Plim\ lowers the \Proto\ probability for a given overdensity, and increases the \ProtoF\ probability.
A liberal choice of both \Plim\ and \Clim\ leads to high protocluster probabilities, but the probability of probing a field region at low overdensity is still high.
Choosing both \Plim\ and \Clim\ conservatively leads to \PProto\ probabilities dominating.
We recommend choosing values of \Plim\ and \Clim\ motivated by the completeness and purity of the protocluster population, given the aperture choice and selection.


\bsp	
\label{lastpage}

\end{document}